\documentclass[]{aa}
\usepackage{graphicx}
\usepackage{natbib}
\bibpunct{(}{)}{;}{z}{}{,}
\def\FIR{\ifmmode {\,$\tau_{\rm FIR}$} \else $\,\tau_{\rm FIR}$\fi}
\def\tk{\ifmmode {\,$T_{\rm k}$} \else $\,T_{\rm k}$\fi}
\def\mic{\ifmmode {\,\mu{\rm m}} \else $\,\mu{\rm m}$\fi}       % micron
\def\kms{\ifmmode {\,{\rm km\,s^{-1}}}                          % km s-1
	\else {\hbox{$\,$ {\rm km$\,$s$^{\rm -1}$}}}\fi}
\def\mo {\ifmmode {\,{\it M}\odot} \else $\,M$\odot\fi} % M solar
\def\lo {\ifmmode {\,{\it L}\odot} \else $\,L$\odot\fi} % L solar
\def\my {\ifmmode {\,{\it M}\solar\,{\rm yr^{-1}}}              % Msol/year
	\else {$\,M$\solar$\,$yr$^{\rm -1}$}\fi}
\def\cmm#1{\ifmmode {\,{\rm cm^{-#1}}}          % cm-1, cm-2, cm-3,....  
	\else \hbox{$\,${\rm cm$^{\rm -#1}$}}\fi}
\chardef\isp="10\def\i{\'\isp}		% dotless i withaccent
\def\as {\ifmmode {^{\scriptscriptstyle\prime\prime}}           % arcsec
	\else $^{\scriptscriptstyle\prime\prime}$\fi}
\def\am {\ifmmode {^{\scriptscriptstyle\prime}}                 % arcmin
	\else $^{\scriptscriptstyle\prime}$\fi}
\def\deg {\ifmmode^\circ\else$^\circ$\fi}                       % degree
\def\raw {\ifmmode\rightarrow\else$\rightarrow$\fi}             % rightarrow
\def\x {\ifmmode\times\else$\times$\fi}                         % times (x)
\def\gsim {\ifmmode {\buildrel>\over\sim}               % greater or similar
	\else {\lower.6ex\hbox{$\buildrel>\over\sim$}}\fi}
\def\lsim {\ifmmode {\buildrel<\over\sim}               % less or similar
	\else {\lower.6ex\hbox{$\buildrel<\over\sim$}}\fi}
\def\ra[#1 #2 #3.#4]{ #1$^{\rm h}$#2$^{\rm m}$#3$^{\rm s}$.#4}  % RA
\def\dec[#1 #2 #3.#4]{ #1\deg#2\am#3{\as}.#4}             % declination
\def\rax[#1 #2 #3]{RA: #1$^{\rm h}$#2$^{\rm m}$#3$^{\rm s}$}%RA with 
\def\decx[#1 #2 #3]{Dec:#1\deg#2\am#3\as}          % Dec with no fraction
\def\h2{\rm H$_2$}                      % molecular hydrogen
\def\apj{{\rm ApJ}}                     % Astrophys. J.
\def\apjs{{\rm ApJS}}                   % Astrophys. J. Supl.
\def\apjl{{\rm ApJ Let}}                % Astrophys. J. (Letters)
\def\mnras{{\rm MNRAS}}                 % Monthly Notices...  
\begin{document}
\title{Chemical evolution in the environment of  
intermediate mass young stellar objects:}
\subtitle{ NGC~7129~--~FIRS~2 and
LkH$\alpha$~234}
  \author{A. Fuente\inst{1}
  \and J.R. Rizzo\inst{1,2} 
  \and P. Caselli\inst{3}
  \and R. Bachiller\inst{1}
 \and C. Henkel\inst{4}}
\institute{Observatorio Astron\'omico Nacional (IGN), Campus
 Universitario, Apdo. 112, E-28800 Alcal\'a de Henares (Madrid), Spain     
 \and
Departamento de F{\i}sica, Universidad Europea de Madrid, Urb.\ El Bosque,
       E-28670 Villaviciosa de Od\'on, Spain
\and
Osservatorio Astrofisico di Arcetri, Largo Enrico Fermi 5, 50125 Firenze, Italy
\and
Max-Planck-Institut f\"ur Radioastronomie, Auf dem H\"ugel 69, 53121 Bonn, Germany} 

\offprints{A. Fuente, \email{a.fuente@oan.es}}
       
%\date{Received July 15/ 1995, accepted }
%
%
\abstract{We have carried out a molecular survey of the Class 0 IM protostar NGC \,7129 -- FIRS 2 
(hereafter FIRS 2) and the Herbig
Be star LkH$\alpha$ 234 with the aim of studying the chemical evolution of the envelopes of 
intermediate-mass (IM) young stellar
objects (YSOs). Both objects have similar luminosities ($\sim$500~L$_\odot$) and
are located in the same molecular cloud which minimizes the 
chemical differences due to different stellar masses or initial cloud
conditions. Moreover, since they are located at the same distance,
we have the same spatial resolution in both objects. A total of 17 molecular species 
(including rarer isotopes) have been observed in both objects and the structure of their
envelopes and outflows is determined with unprecedent detail. 

Our results show that the protostellar envelopes are dispersed and warmed up during the 
evolution to become a pre-main sequence star. In fact, the envelope mass decreases by 
a factor $>$5 from FIRS 2 to LkH$\alpha$~234, 
while the kinetic temperature increases  
from $\sim$13~K to 28~K.  On the other hand, there is no molecular outflow 
associated with LkH$\alpha$~234. The molecular outflow seems to stop before the star becomes
visible. 
  
These physical changes strongly affect the chemistry
of their envelopes. The N$_2$H$^+$ and NH$_3$ abundances seem to be quite
similar in both objects. However, the H$^{13}$CO$^+$ abundance is a factor of $\sim$3 lower
in the densest part of FIRS 2 than in LkH$\alpha$ 234, very likely because of depletion. In contrast, the
SiO abundance is larger by a factor of $\sim$100 in FIRS~2 than in LkH$\alpha$~234. 
CS presents a complex behavior since its emission arises in different envelope
components (outflow, cold envelope, hot core) and could also suffer from depletion. 
The CH$_3$OH and H$_2$CO column densities are very similar in FIRS~2 and
LkH$\alpha$~234  which implies that the beam averaged abundances are
a factor $>$5 larger in LkH$\alpha$~234 than in FIRS~2. 
The same case is found for the
PDR tracers CN and HCN which have similar column densities in both objects. Finally,
a complex behavior is found for the deuterated compounds. While the DCO$^+$/H$^{13}$CO$^+$
ratio decreases by a factor of $\sim$4 from FIRS~2 to LkH$\alpha$~234, the D$_2$CO/H$_2$CO
ratios is within a factor 1.5 in both objects. 
The detection of a warn CH$_3$CN component with T$_k$$>$63~K shows the existence of a hot core in FIRS~2.
Thus far, only a handful of hot cores have been detected
in low and intermediate mass stars.

Based on our results in FIRS~2 and LkH$\alpha$ 234, we propose some abundance
ratios that can be used as chemical clocks for the envelopes of IM YSOs. 
The SiO/CS, CN/N$_2$H$^+$, HCN/N$_2$H$^+$, DCO$^+$/HCO$^+$ and
D$_2$CO/DCO$^+$ ratios are good diagnostics of the protostellar evolutionary stage.  

}
\maketitle

\setlength\unitlength{1cm}
\begin{figure*}
\vspace{12cm}
\includegraphics{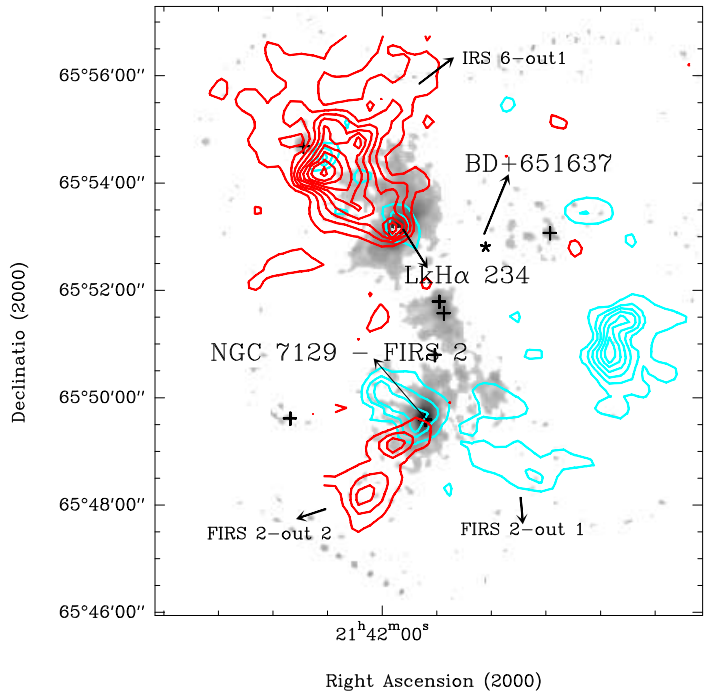}
%\end{picture}
\caption{Map of the 1.3mm continuum flux toward NGC 7129. 
Levels are 22.3, 44.6, 89.2, 178.4 to
713.6 mJy/beam by 178.4 mJy/beam. Crosses indicate the millimeter sources and
stars the infrared sources. The red and blue contours represent 
the redshifted and blueshifted high-velocity gas of the nebula as
traced by the $^{12}$CO J=2$\rightarrow$1 line. 
Velocity intervals are [-30,-13] kms$^{-1}$
for the blue lobe (grey contours) and  [-7,11] kms$^{-1}$
for the red-lobe (dark contours). Contours are 10 to 60 K kms$^{-1}$ in steps 
of 10 K kms$^{-1}$ for the blue lobe and 10 to 100 K kms$^{-1}$ in steps 
of 15 K kms$^{-1}$ for the red lobe. }
\end{figure*}

\section{Introduction}

Chemistry is a powerful tool to study young stellar objects (YSOs) and their
environments.
On the one hand, chemistry is a diagnostic tool of the different
envelope components. On the other hand, it is a good  time indicator during the protostellar
evolution. Chemical studies have been used to determine the physical structure of low-mass
YSOs \citep[e.g.][]{mar04,jor04a,jor04b}. These studies have
revealed, for example, the presence of warm regions where the ices evaporate giving
rise to regions similar to hot cores in massive protostars. However, the chemistry of the
two classes of objects is different and rises questions on the mechanisms that
lead to the observed chemical complexity and its dependence on the
stellar mass \citep{caz03,bot04}.

Chemistry has also been used
as a time evolution indicator in both low-mass and high-mass objects.
For instance, \citet{mar04} studied a sample of low-mass Class 0 objects and found
some indications that the H$_2$CO abundance increases during the protostellar
evolution. However, the different abundances can be due 
to different initial cloud conditions from which the protostars evolved. 
Further chemical studies are required to clarify this subject.

In this paper, we present a chemical study of the two intermediate mass 
(IM) YSOs,  FIRS 2 and LkH$\alpha$ 234. Contrary to low-mass
protostars, IM YSOs have been very little studied thus far, specially young protostars.
For example, out of 42 Class 0 sources  compiled by \citet{and00}, 
only six had luminosities in excess of 40 L$_{\odot}$ (the precursors to HAe stars) 
and only one had a luminosity of $\sim 10^3$ L$_{\odot}$ (the precursor to HBe stars).  
IM YSOs (M$_*$ $\sim$ 2 -10 M$_\odot$) are not only an important link between low 
mass and high mass stars but they share many
characterisitcs of high mass star formation (clustering, PDR/HII regions) without the disadvantage
of being too distant (d $\lsim$ 1 Kpc), or too complex.
They are also important for the understanding of planet 
formation since Herbig Ae stars are the precursors of Vega-type systems. 
On a larger scale, they dominate the mean UV interstellar field 
in our Galaxy \citep{wol03}.

\section {Target selection and observational strategy}

FIRS~2 has been classified
as a Class 0 IM object \citep{eir98} and  is, very likely, the
youngest IM object known at present. An energetic bipolar molecular outflow is
associated with it \citep{fue01}. LkH$\alpha$ 234 is a embedded HBe star
which still keeps a massive envelope (M$\sim$16 M$_{\odot}$)
but no bipolar molecular outflow seems to be driven by it \citep{fue01,fue02}.  
These objects have the peculiarity of having similar luminosities ($\sim$ 500 L$_{\odot}$) and 
be located in the same molecular cloud. 
During the protostellar and pre-main sequence evolution, the luminosity
remains quite constant for  a given stellar mass \citep{and00}. Thus, the same luminosity
implies similar stellar mass.
In addition, both sources are located in the same molecular cloud.
This minimizes chemical effects due to 
very different stellar mass (both of them have M$_*$ $\sim$ 5 M$_\odot$) 
and/or different initial cloud conditions. Thus, in spite of our reduced sample, we have
an excellent opportunity to find an evolutionary track for IM YSOs. 
In Fig. 1 we show the 1.3mm continuum map of the
reflection nebula NGC~7129. Both, FIRS~2 (NGC~7129--FIRS~2) and LkH$\alpha$~234
(NGC~7129 -- FIRS~1) are associated with intense centrally 
peaked 1.3mm continuum and far infrared sources. We have also overlaid the red and
blue lobes (as traced by the $^{12}$CO J=2$\rightarrow$1 line) of the bipolar outflows
associated to these YSOs (see Fuente et al. 2001 for a more detailed description).
Clustering becomes significant in this range of stellar masses \citep{tes99}.
Interferometric observations towards LkH$\alpha$ 234 show the existence of a young 
infrared companion (IRS 6) which is very likely the exciting source of the bipolar 
molecular outflow and the [SII] jet detected by \citet{ray90}.
The quadrupolar morphology of the ouflow detected in  NGC~7129--FIRS~2 is also due to the 
superposition of two bipolar molecular outflows FIRS 2-out 1 and FIRS 2-out 2 \citep{fue01}.
\citet{fue01} proposed that FIRS 2-out 1 is associated with 
the Class 0 protostar while FIRS 2-out 2 is more likely associated with a more evolved infrared star
(FIRS~2~-~IR). 

A complex chemical evolution
occured in the YSO envelopes during the protostellar evolution.
This involves accretion
of species in an icy mantle during the pre-collapse phase, followed
by grain-surface chemistry and evaporation of ices once the YSO has
started to heat its surroundings \citep[e.g.][]{bro88}. 
In massive YSOs, the evaporated molecules drive a
rapid high-temperature chemistry for a period of  10$^4$--10$^5$
years, resulting in the complex, saturated organic molecules
(CH$_3$OCH$_3$,CH$_3$CN,C$_2$H$_5$OH,...), which are characteristic of
a hot core \citep[e.g.][]{cha92,cas93,rod03,nom04,vit04}.  
Once most of the envelope has been
cleared up, the UV radiation can escape to form a
photon-dominated-region (PDR) and, in the case of massive stars, an HII
region. Simultaneously, the energetic bipolar outflows develope a shock chemistry
in the surrounding molecular cloud.

To discern between the different envelope components and determine the protostellar
envelope evolution,
we have selected a set of molecular tracers. Specifically,
NH$_3$, N$_2$H$^+$, H$^{13}$CO$^+$ and HC$^{18}$O$^+$ have been 
observed as tracers of  the {\it cold envelope}; 
the volatile species CH$_3$OH and H$_2$CO to study the {\it warm envelope} where the
icy grain mantles have been evaporated; the complex molecule CH$_3$CN to trace
the {\it hot core};  and, CN and HCN in order to trace the incipient  {\it PDR}.
In addition, we have also observed several CS and C$^{34}$S lines that are useful
to constrain the physical conditions of the envelope and
SiO as an excellent tracer of the shock chemistry
associated to the {\it outflow}. The observation of the
deuterated compounds N$_2$D$^+$,DCO$^+$ and D$_2$CO 
also provide us some information about
how the deuterium fractionation is affected by the YSO 
evolution. 
Obviously, the $``$molecular tracer$"$--$``$envelope component$"$ correspondence is
not unique and all the species have contributions from other envelope components.
Arguments based on the morphology of the emission, kinematics and different
excitation conditions are used to discern between the different components in these cases.
				
\section{Observations}

The (J,K) = (1,1), (2,2), (3,3) and (4,4) inversion lines of ammonia were 
observed using the Effelsberg 100--m radiotelescope of the MPIfR in December 
2000 and October 2002. We observed only one position in both sources.
The half power beam width (HPBW) of the telescope at the rest frequency of the NH$_3$
lines, 23.7 GHz, was 40\arcsec. We used the new cooled dual-channel HEMT
K--band receiver with a typical system temperature of 200\,K on a main beam
brightness temperature scale. The 8192--channel autocorrelator was used as the
backend. The four ammonia lines were observed simultaneously with a total
bandwidth of 10 MHz and a channel separation of 0.098 \kms. 
We estimate that line intensities are accurate to within $\pm$15\%.

The main set of observations were carried out with
the IRAM 30m telescope in Pico de Veleta (Spain)
during three different observing periods in May 1999, 
July 2002 and August 2003. The list of observed lines,
the telescope characteristics at each frequency and a
summary of the observations are shown 
in Table 1. When possible, all the lines of the same molecule
were observed simultaneously  in order to avoid observational errors.
The backends were an autocorrelator split in several parts and 
a 256 $\times$ 100 kHz filter-bank. All the lines were observed
with a spectral resolution of $\sim$ 78 kHz except in the cases that
are explicitly indicated in Table 1. The intensity scale used in this paper
is main brightness temperature. Comparing the intensity of some pattern lines in
different observing periods, we estimate that the calibration
is accurate within a 20 \% at 3mm and within a 30\% at 1.3mm.
We have observed a high S/N ratio spectra towards the star positions 
for all the lines listed in Table 1. In addition, we have carried out small maps 
in the most intense lines (see Table 1).  In the case of the 
SiO 2$\rightarrow$1, SiO 3$\rightarrow$2,
H$^{13}$CO$^+$ 1$\rightarrow$0 and CS 3$\rightarrow$2 
lines we have mapped the entire outflows as traced by the high velocity
CO emission. Fits to the oberved lines at the (0,0) position are shown in 
Tables 2 and 3.

\begin{table*}[t]
\caption{Description of the IRAM 30m observations}
\begin{tabular}{lr cccc} \\  \hline
\multicolumn{1}{c}{Line} &       
\multicolumn{1}{c}{Freq. (GHz)} &
\multicolumn{1}{c}{HPBW} &
\multicolumn{1}{c}{$\eta_{MB}$} &
 \multicolumn{1}{c}{Observed positions}  \\ \hline
N$_2$H$^+$ 1$\rightarrow$0   &   93173.2  &   26.5$''$ &  0.71  & 60$''$$\times$60$''$ map\\ 
CN 1$\rightarrow$0 & 113490.0 & 23$''$ & 0.65 &   60$''$$\times$60$''$ map (FIRS 2), 
60$"$$\times$100$''$ map (LkH$\alpha$ 234) \\
HCN 1$\rightarrow$0 & 88631.8 &  27.5$''$ &  0.73 &  60$''$$\times$60$''$ map (FIRS 2), 
60$"$$\times$100$''$ map (LkH$\alpha$ 234) \\
%NH$_2$D (1,1)         & 85926.0    & 27.5$''$ & 0.73 &   IRAM  & map \\ 
H$^{13}$CO$^+$     1$\rightarrow$0 & 86754.3   & 27.5$''$ & 0.73  & complete map\\
HC$^{18}$O$^+$     1$\rightarrow$0 & 85162.2  &  27.5$''$ &  0.73 &   (0,0) \\ 
SiO 2$\rightarrow$1  &  86846.9 & 27.5$''$ & 0.73 &   complete map \\
SiO 3$\rightarrow$2   &   130268.6  & 20$''$ & 0.60  &  complete map \\ 
CS  3$\rightarrow$2    & 146969.0   &  16.5$''$ & 0.55 &   complete map\\
C$^{34}$S 2$\rightarrow$1 &  96412.9 & 25.9$''$ & 0.70  &  (0,0) \\
C$^{34}$S 3$\rightarrow$2 &  144617.1 & 17$''$  &  0.55  &  (0,0) \\
C$^{34}$S 5$\rightarrow$4 &  192818.5 & 13$''$   & 0.47 &   (0,0) \\ 
CH$_3$OH  2(1,3)$\rightarrow$1(1,3) & 96755.5   &  26$''$ &  0.70 & 72$''$$\times$72$''$ map \\
CH$_3$OH  2(0,3)$\rightarrow$1(0,3) & 96744.6   &  26$''$ & 0.70  & 72$''$$\times$72$''$ map\\
CH$_3$OH  2(1,4)$\rightarrow$1(1,4) & 96741.4   &  26$''$ & 0.70  & 72$''$$\times$72$''$ map\\
CH$_3$OH  2(0,1)$\rightarrow$1(0,1) & 96739.4   &  26$''$ & 0.70  & 72$''$$\times$72$''$ map\\
CH$_3$OH  3(0,3)$\rightarrow$2(0,3) & 145093.7 &  17$''$  & 0.55  & (0,0) \\
CH$_3$OH  3(1,4)$\rightarrow$2(1,4) & 145097.5 &  17$''$ & 0.55  & (0,0) \\
CH$_3$OH  3(0,1)$\rightarrow$2(0,1) & 145103.2 &  17$''$  & 0.55 & (0,0) \\
CH$_3$OH  3(2,2)$\rightarrow$2(2,2)$^a$ & 145124.4 &  17$''$ & 0.55 & (0,0) \\
CH$_3$OH  3(1,3)$\rightarrow$2(1,3) & 145131.9  &  17$''$ & 0.55  & (0,0) \\
CH$_3$OH  5(2,4)$\rightarrow$4(2,4) & 241904.1  &  10$''$  & 0.40   & 72$''$$\times$72$''$ map\\
CH$_3$OH  5(0,1)$\rightarrow$4(0,1) & 241791.4  &  10$''$  & 0.40 & 72$''$$\times$72$''$ map\\
CH$_3$OH  5(1,4)$\rightarrow$4(1,4) & 241767.2  &  10$''$   & 0.40 & 72$''$$\times$72$''$ map\\
CH$_3$OH  5(3,1)$\rightarrow$4(3,1) & 241832.9$^a$  &  10$''$  & 0.40   & 72$''$$\times$72$''$ map\\
CH$_3$OH  5(2,2)$\rightarrow$4(2,2) & 241842.3  &  10$''$  & 0.40  & 72$''$$\times$72$''$ map\\
CH$_3$OH  8(1,4)$\rightarrow$7(0,3) & 229758.7  &  10.5$''$  & 0.42  & (0,0) \\ 
CH$_3$OH  11(1,4)$\rightarrow$10(2,4)                 & 104300.5   &  24.5$''$ & 0.68  & (0,0) \\ 
H$_2$CO               3$_{12}$$\rightarrow$2$_{11}$ & 225697.8   &  10.5$"$ &  0.43 & 60$''$$\times$60$''$ map (FIRS 2),  
72$''$$\times$72$''$ map (LkH$\alpha$ 234) \\ 
H$_2$CO               2$_{12}$$\rightarrow$1$_{11}$ & 140839.5   &  18$''$ & 0.57  & 60$''$$\times$60$''$ map (FIRS 2),
72$''$$\times$72$''$ map (LkH$\alpha$ 234)\\
H$_2$$^{13}$CO    2$_{12}$$\rightarrow$1$_{11}$  &  137449.9 &   18$''$ & 0.57 & (0,0) \\ 
CH$_3$CN 5(0)$\rightarrow$4(0)  & 91987.0  & 27$''$ &  0.72    & (0,0) \\
CH$_3$CN 5(1)$\rightarrow$4(1)  & 91985.3  & 27$''$ &  0.72    & (0,0) \\
CH$_3$CN 5(2)$\rightarrow$4(2)  & 91980.0  & 27$''$ &  0.72 & (0,0) \\
CH$_3$CN 5(3)$\rightarrow$4(3)  & 91971.4  & 27$''$ & 0.72 & (0,0)  \\
CN 2$\rightarrow$1 & 226874.7 & 11$''$  & 0.42  & 60$''$$\times$60$''$ map (FIRS 2), 60$"$$\times$100$''$
(LkH$\alpha$ 234) \\ 
N$_2$D$^+$           3$\rightarrow$2 & 231321.7 &  10.5$''$ & 0.42 & (0,0) \\ 
DCO$^+$               2$\rightarrow$1  & 144077.3 & 17$''$ &  0.56  & 60$''$$\times$60$''$ map\\
DCO$^+$               3$\rightarrow$2  & 216112.6 &   11$''$ & 0.44  & 60$''$$\times$60$''$ map \\ 
D$_2$CO              4$_{04}$$\rightarrow$3$_{03}$ & 231410.3  &  10.5$''$ & 0.42 & (0,0)   \\ 
\hline
\end{tabular}

\noindent
$^a$ Only observed with a frequency resolution of 1 MHz.
\end{table*}

\begin{table*}[t]
\caption{Observational parameters towards NGC 7129 -- FIRS 2}
{\scriptsize
\begin{tabular}{ll ccc} \\ 
%\multicolumn{6}{c}{NGC 7129 -- FIRS 2} \\
 \hline
\multicolumn{1}{l}{Line} &       
\multicolumn{1}{c}{T$_{MB}$$\times$$\tau_m$} &
\multicolumn{1}{c}{V (km s$^{-1}$)} &
\multicolumn{1}{c}{$\Delta$V (km s$^{-1}$)} &
\multicolumn{1}{c}{$\tau_m$ } \\ \hline
NH$_3$ (1,1) & 3.08(0.07) & -9.59(0.01) & 1.0(0.1) & 0.1$^a$ \\ 
NH$_3$ (2,2) & 0.70(0.02) & -9.63(0.02) & 1.3(0.1) & 0.1$^a$ \\
NH$_3$ (3,3) & 0.12(0.01) & -9.35(0.18) & 3.8(0.5) & 0.1$^a$ \\
NH$_3$ (4,4) & \multicolumn{4}{c}{rms = 0.05 K kms$^{-1}$ with $\Delta$v=3.0 kms$^{-1}·$} \\ 
N$_2$H$^+$ 1$\rightarrow$0    &  9.22(0.02) & -9.51(0.01) & 1.91(0.01) & 1.28(0.10)    \\
%(--32$''$,0)                                   &  1.18(0.08)  & -9.83(0.06) & 1.83(0.12) & 0.10(0.72) \\
CN 1$\rightarrow$0 & 2.3(0.3) & -9.7(0.1) & 1.4(0.1) & 1.9 \\
HCN 1$\rightarrow$0  & 2.3(0.3)  & -9.3(0.1) & 3.1(0.1) & 0.7 \\
%NH$_2$D (1,1)                     & 85.926       &  0.55(0.01) & -9.80(0.01) & 1.32(0.02) & 0.61(0.06)   \\ 
\hline
\multicolumn{1}{l}{Line} &       
\multicolumn{1}{c}{Area (K kms$^{-1}$)} &
\multicolumn{1}{c}{V (km s$^{-1}$)} &
\multicolumn{1}{c}{$\Delta$V (km s$^{-1}$)} &
\multicolumn{1}{c}{T$_{MB}$ (K)}  \\ \hline
H$^{13}$CO$^+$     1$\rightarrow$0    & 1.92(0.02) & -9.74(0.01) & 1.27(0.02) & 1.39  \\
%(-32$''$,0)                                         & 0.57(0.06)  & -9.54(0.06) & 1.15(0.17) & 0.46 \\
HC$^{18}$O$^+$     1$\rightarrow$0   & 0.17(0.02) & -9.68(0.09) & 1.1(0.2)     & 0.14 \\ 
SiO 2$\rightarrow$1                          &  1.55(0.09) & -7.5(0.2) & 6.8(0.4)  & 0.21 \\
                                                      &   0.23(0.06) & -9.5(0.1) & 1.6(0.4)  & 0.13 \\
SiO 3$\rightarrow$2     &  2.50(0.20) & -7.3(0.2) & 6.9(0.7)  & 0.34 \\ 
CS  3$\rightarrow$2     &  3.69(0.12)   & -9.1(0.1) & 5.81(0.24)   &  0.60 \\
                                 &  1.88(0.09)   &   -9.9(0.1) &  1.19(0.04)  & 1.48   \\
C$^{34}$S 2$\rightarrow$1  & 0.16(0.08) & -7.8(0.8) & 3.1(1.3)  & 0.05 \\
                                        & 0.28(0.07) & -9.8(0.1) & 1.3(0.2)  & 0.20 \\
C$^{34}$S 3$\rightarrow$2  & 1.05(0.08) & -9.4(0.1) & 4.3(0.4)   & 0.23 \\
C$^{34}$S 5$\rightarrow$4   & 1.21(0.16) & -8.8(0.3) & 4.2(0.7)  & 0.27 \\
CH$_3$OH  2(1,3)$\rightarrow$1(1,3)  & 0.6(0.3)         & -9(1)             &  5(3) &  0.11 \\
CH$_3$OH  2(0,3)$\rightarrow$1(0,3)  & 1.85(0.06)     & -8.20(0.09)     &  6.1(0.2) & 0.28  \\
CH$_3$OH  2(1,4)$\rightarrow$1(1,4)  & 4.71(0.05)     & -9.2(0.2)        &  3.0(0.2)  &  1.47 \\
CH$_3$OH  2(0,1)$\rightarrow$1(0,1)  & 5.83(0.07)     & -9.58(0.01)     & 5.3(0.1)  &  1.03 \\
CH$_3$OH  3(0,3)$\rightarrow$2(0,3)  & 4.12 (0.04)    & -8.78(0.08)     & 6.98(0.08) & 0.55  \\
CH$_3$OH  3(1,4)$\rightarrow$2(1,4)  &  8.9(0.09)     & -8.32(0.01)     & 5.71(0.07)  & 1.46 \\
CH$_3$OH  3(0,1)$\rightarrow$2(0,1) &   9.6(0.1)       & -8.34(0.02)     & 5.07(0.06) &  1.78 \\
CH$_3$OH  3(2,2)$\rightarrow$2(2,2) &  2(1)      &  -12(4)          & 7(10)  & 0.27 \\
CH$_3$OH  3(1,3)$\rightarrow$2(1,3) & 1.9(0.1)  & -8.7(0.2)       & 6.4(0.5) & 0.28 \\
CH$_3$OH  5(2,4)$\rightarrow$4(2,4) & 3(1)       & -8.9(0.7)        & 4(2)  &  0.70 \\
CH$_3$OH  5(0,1)$\rightarrow$4(0,1) & 12.3(0.7) & -7.4(0.1)     & 4.6(0.3) & 2.51 \\
CH$_3$OH  5(1,4)$\rightarrow$4(1,4) &  8.8(0.8)  & -7.8(0.2)    & 4.6(0.5)  & 1.80 \\
CH$_3$OH  5(3,1)$\rightarrow$4(3,1) & 1.7(1.7)   & -8.1(3)      & 6.9(7)   &  0.23 \\
CH$_3$OH  5(2,2)$\rightarrow$4(2,2)  & 1.3(1.5)   & -10(4)       & 7(11)   & 0.17 \\
CH$_3$OH  8(1,4)$\rightarrow$7(0,3)  &  4.5(0.4)  & -7.7(0.1)    & 4.3(0.3) & 0.98 \\ 
CH$_3$OH  11(1,4)$\rightarrow$10(2,4)  & 0.14(0.04) & -9.9(0.4)   & 3.6(5.9) & 0.04 \\ 
H$_2$CO     3$_{12}$$\rightarrow$2$_{11}$ & 2.3(0.2) & -9.6(0.1) & 1.3(0.1) & 1.66   \\
                                                               & 6.7(0.3) & -8.6(0.2)     & 7.7(0.5) & 0.81   \\
H$_2$CO               2$_{12}$$\rightarrow$1$_{11}$  & 3.0(0.1) & -9.66(0.01) & 1.41(0.03) & 2.00   \\
                                                                         & 6.0(0.1) & -8.97(0.06) & 6.3(0.2)    & 0.91    \\
H$_2$$^{13}$CO    2$_{12}$$\rightarrow$1$_{11}$  & 0.33(0.04) & -9.4(0.2) & 2.8(0.5) & 0.11  \\ 
CH$_3$CN 5(0)$\rightarrow$4(0)   & 0.24(0.05) & -9.1(0.4) & 4.2(0.8) & 0.05 \\
CH$_3$CN 5(1)$\rightarrow$4(1)   & 0.25(0.05) & -9.1(0.6) & 5(1)       &  0.05 \\
CH$_3$CN 5(2)$\rightarrow$4(2)   & 0.11(0.02) & -9.7(0.9) & 4(2)        & 0.02 \\
CH$_3$CN 5(3)$\rightarrow$4(3)   & 0.12(0.02) & -8(1) & 7(3)             & 0.02 \\
CN 2$\rightarrow$1  & 2.74(0.24) & -9.7(0.1) & 2.6(0.2) & 1.0 \\
N$_2$D$^+$           3$\rightarrow$2 & 0.64(0.05) & -9.90(0.04) & 1.1(0.1)      & 0.53   \\
DCO$^+$               2$\rightarrow$1  & 2.93(0.04) & -9.86(0.01) & 1.0(0.1)      & 2.61    \\
DCO$^+$               3$\rightarrow$2  & 1.9(0.1)     & -9.73(0.02) & 1.0(0.1)      & 1.83   \\
D$_2$CO              4$_{04}$$\rightarrow$3$_{03}$   & 0.3(0.1)    & -9.6(0.3) & 3.9(1)    & 0.05  \\
\hline
\end{tabular}

\noindent
$^a$ In case of optically thin emission, $\tau_m$ cannot be determined and is set arbitrarily to 0.1.
}
\end{table*}

\begin{table*}[t]
\caption{Observational parameters towards LkH$\alpha$ 234}
{\scriptsize
\begin{tabular}{llccc}\\
%\multicolumn{6}{c}{LkH$\alpha$ 234} \\ 
\hline
\multicolumn{1}{l}{Line} &       
\multicolumn{1}{c}{T$_{MB}$$\times$$\tau_m$} &
\multicolumn{1}{c}{V (km s$^{-1}$)} &
\multicolumn{1}{c}{$\Delta$V (km s$^{-1}$)} &
\multicolumn{1}{c}{$\tau_m$ } \\ \hline
NH$_3$ (1,1) & 0.60(0.03) & -9.87(0.02) & 0.9(0.2) & 0.86(0.07) \\ 
NH$_3$ (2,2) & 0.27(0.04) & -9.94(0.05) & 1.1(0.2) & 0.1$^a$ \\
NH$_3$ (3,3) & 0.10(0.03) & -11(1)        & 1.6(0.7) & 0.1$^a$ \\
NH$_3$ (4,4) & \multicolumn{4}{c}{rms = 0.13 K kms$^{-1}$ with $\Delta$v=1.1 kms$^{-1}·$} \\ 
N$_2$H$^+$ 1$\rightarrow$0    &  1.53(0.06)  & -9.48(0.01) & 1.82(0.04) & 1.32(0.7)  \\
CN 1$\rightarrow$0  & 1.9(0.2) & -9.9(0.1) & 2.6(0.1) & 0.8 \\
HCN 1$\rightarrow$0 & 3.2(0.3)  & -10.0(0.1) & 2.9(0.1) & 0.1$^a$ \\  
%NH$_2$D (1,1)          & 0.10(0.02) & -9.85(0.05) & 0.91(0.08) & 0.97(0.60)  \\
\hline
\multicolumn{1}{l}{Line} &       
\multicolumn{1}{c}{Area (K kms$^{-1}$)} &
\multicolumn{1}{c}{V (km s$^{-1}$)} &
\multicolumn{1}{c}{$\Delta$V (km s$^{-1}$)} &
\multicolumn{1}{c}{T$_{MB}$ (K)}  \\ \hline 
H$^{13}$CO$^+$     1$\rightarrow$0  &  1.00(0.03) & -9.80(0.03) & 1.79(0.07) & 0.52 \\
HC$^{18}$O$^+$     1$\rightarrow$0  &  0.14(0.04) & -10.1(0.2) & 2(1) & 0.06 \\ 
SiO 2$\rightarrow$1  & 0.31(0.06) & -9.9(0.9)  & 8.5(2.0)  & 0.03  \\
SiO 3$\rightarrow$2  & 0.60(0.08)  & -8.7(0.2) & 3.2(0.5)  &  0.18  \\ 
CS 3$\rightarrow$2    & 14.9(0.2) & -9.7(0.1)  & 3.0(0.1)   &  4.67  \\
C$^{34}$S 2$\rightarrow$1  & 0.66(0.04) & -9.9(0.1) & 3.4(0.2)  &  0.18  \\
C$^{34}$S 3$\rightarrow$2  & 1.41(0.05) & -9.5(0.1) & 3.2(0.2)   & 0.41 \\
C$^{34}$S 5$\rightarrow$4   & 1.71(0.22) & -9.1(0.3) & 4.5(0.8)   &  0.36 \\ 
CH$_3$OH 2(1,3)$\rightarrow$1(1,3)  & 0.3(0.1)   & -7(3)      &  12(5)     &  0.02 \\
CH$_3$OH 2(0,3)$\rightarrow$1(0,3) & 0.20(0.04) & -9.8(0.3) & 2.9(0.7)   &  0.06 \\
CH$_3$OH 2(1,4)$\rightarrow$1(1,4) & 0.48(0.03) & -9.6(0.1) & 2.2(0.2)   &  0.20 \\
CH$_3$OH 2(0,1)$\rightarrow$1(0,1) & 0.64(0.04) & -9.5(0.1) & 5.2(0.4)    & 0.11 \\
CH$_3$OH 3(0,3)$\rightarrow$2(0,3) & 1.27(0.09)  & -9.5(0.1) & 4.6(0.4)  &  0.26 \\
CH$_3$OH 3(1,4)$\rightarrow$2(1,4) &  1.91(0.09)  & -9.2(0.1) & 4.8(0.3)  &  0.37 \\
CH$_3$OH 3(0,1)$\rightarrow$2(0,1) & 1.94(0.07)  & -9.2(0.1) & 3.3(0.2)  &  0.55 \\
CH$_3$OH 3(2,2)$\rightarrow$2(2,2) &  1.6(0.4)   &  -12(1)  & 9(4)   &  0.17 \\
CH$_3$OH 3(1,3)$\rightarrow$2(1,3) &  1.3(0.4)   & -10(1) & 8(3)     & 0.15 \\
CH$_3$OH 5(2,4)$\rightarrow$4(2,4) &  2.6(0.5)   & -9.4(0.5) & 5(1) & 0.49  \\
CH$_3$OH 5(0,1)$\rightarrow$4(0,1) & 2.9(0.2)  & -8.8(0.1) & 4.1(0.6)     &  0.66 \\
CH$_3$OH 5(1,4)$\rightarrow$4(1,4) &  2.1(0.2) & -9.2(0.1) & 3.1(0.2)     &  0.64 \\
CH$_3$OH 8(1,4)$\rightarrow$7(0,3) &  2.0(0.3)      &  -9.7(0.5) & 7(1) & 0.27  \\
CH$_3$OH 11(1,4)$\rightarrow$10(2,4)  &  0.13(0.04)  &  -9.9(0.8) & 6(2)  &  0.02 \\ 
H$_2$CO               3$_{12}$$\rightarrow$2$_{11}$ & 5.4(0.3)  & -9.4(0.1) & 2.2(0.1) & 2.2 \\
                                                                         & 5.9(0.3)   &  -9.8(0.1)  & 5.5(0.2) & 1.0  \\
H$_2$CO               2$_{12}$$\rightarrow$1$_{11}$ & 8.1(0.6)   &  -9.7(0.1)   & 2.4(0.1) & 3.1  \\
                                                                         & 3.1(0.6)    & -9.5(0.2)   &  6(1)     & 0.5  \\
H$_2$$^{13}$CO    2$_{12}$$\rightarrow$1$_{11}$  & 0.24(0.05) & -9.4(0.4)   &  4(1)     & 0.05  \\ 
CH$_3$CN 5(0)$\rightarrow$4(0)  & 0.17(0.02) & -10.2(0.2) & 2.6(0.4) & 0.06 \\
CH$_3$CN 5(1)$\rightarrow$4(1)  & 0.16(0.02) & -9.9(0.2) & 2.9(0.4) & 0.05 \\
CH$_3$CN 5(2)$\rightarrow$4(2)  &  \multicolumn{4}{c}{$<$ 0.07 K kms$^{-1}$
 with $\Delta$v=3.2 km s$^{-1}$} \\
CH$_3$CN 5(3)$\rightarrow$4(3)  & \multicolumn{4}{c}{$<$ 0.07 K kms$^{-1}$
 with $\Delta$v=3.2 km s$^{-1}$} \\
CN 2$\rightarrow$1 & 6.0(0.3) & -9.0(0.1) & 2.9(0.2) & 1.9 \\
N$_2$D$^+$           3$\rightarrow$2  &  \multicolumn{4}{c}{$\sigma$=0.11 K kms$^{-1}$ en $\Delta$v= 1.5 km s$^{-1}$} \\
DCO$^+$               2$\rightarrow$1  &  0.62(0.04) & -9.58(0.04) & 1.5(0.1) & 0.38 \\
DCO$^+$               3$\rightarrow$2  &  0.76(0.07) &  -9.4(0.1)    & 2.2(0.2) & 0.33 \\
D$_2$CO              4$_{04}$$\rightarrow$3$_{03}$ & 0.60(0.2)  &  -9.1(0.5)  &   4(1)     & 0.15  \\ \hline
\hline
\end{tabular}

\noindent
$^a$ In case of optically thin emission, $\tau_m$ cannot be determined and is set arbitrarily to 0.1.
}
\end{table*}

\begin{table*}[t]
\caption{LTE column densities}
\begin{tabular}{llccc|llccc}
\\ \hline
\multicolumn{5}{c|}{NGC 7129 -- FIRS 2} &
\multicolumn{5}{c}{LkH$\alpha$ 234} \\
\multicolumn{1}{c}{Molecule} &
\multicolumn{1}{c}{Comp} &
\multicolumn{1}{c}{T$_{rot}$ (K)} &
\multicolumn{1}{c}{N$^b$ (cm$^{-2}$)}  &
\multicolumn{1}{c|}{$\Omega_s$}  &
\multicolumn{1}{c}{Molecule} &
\multicolumn{1}{c}{Comp} &
\multicolumn{1}{c}{T$_{rot}$ (K)} &
\multicolumn{1}{c}{N$^b$ (cm$^{-2}$)}  &
\multicolumn{1}{c}{$\Omega_s$}  \\ \hline
NH$_3$          &   Cold  &  13             & 4.9 10$^{14}$              &    & 
NH$_3$          &   Cold  &  22             & 4.0 10$^{13}$              & 8$''$ \\
                      &  Warm  &  31 -- 87     & $\sim$ 1.5 10$^{13}$    &     & 
                      &  Warm  &  49 -- 134   & $\sim$ 1.1 10$^{13}$    &        \\ 
N$_2$H$^+$          &       &  13$^a$  &  3.8 10$^{13}$   &     $\sim$21$''$   &
N$_2$H$^+$          &        & 22$^a$  & 1.0 10$^{13}$    &        $\sim$8$''$    \\
H$^{13}$CO$^+$    &       &  13$^a$  &  2.2 10$^{12}$   &              & 
H$^{13}$CO$^+$    &       &  22$^a$  & 1.7 10$^{12}$    &                       \\
CH$_3$OH            &                     &  17         &    7.8 10$^{14}$            &       &
CH$_3$OH$^p$       &                &        24           &       1.4 10$^{14}$  &   \\
                              &  Hot core   &    $>$80          &        $\sim$ 2.0 10$^{14}$    &  &
                              &               &   $>$250           &       $\sim$ 4.0 10$^{14}$   &   \\
                              &                  &                       &                                      & & 
 CH$_3$OH$^e$      &                &  59                   &             4.8 10$^{14}$    &  \\  
\multicolumn{5}{c|}{} & \multicolumn{5}{c}{} \\
H$_2$CO              & Narrow      & $\sim$ 10     & 2.2 10$^{13}$               &  &
H$_2$CO               &          & $\sim$ 11       & 8.0 10$^{13}$ &   \\
                            & Wide      & $\sim$ 9      & 4.8 10$^{13}$               &   &
                            &          &                   &                                   &  \\
%                            &             &                       &                       &        &
%                            & N+W     & $\sim$ 11         & 8.0 10$^{13}$   &          \\
H$_2$$^{13}$CO  &     & $\sim$ 10$^a$  & 2.3 10$^{12}$          &   & 
H$_2$$^{13}$CO  &     & $\sim$ 11$^a$      & 2.4 10$^{12}$  &  \\ 
CH$_3$CN          &   Hot core       &     63              & 3.6 10$^{12}$          &     &
CH$_3$CN          &          &  $<$53           & $>$ 2.0 10$^{12}$$^b$  & \\
CN                     &             &     5               &  4.8 10$^{13}$           &   &
CN                     &          &     6.5             &  6.1 10$^{13}$           &    \\
HCN                   &          &    5$^a$        & 1.6 10$^{13}$  &     &
HCN                   &           & 6.5$^a$       &  2.1 10$^{13}$  &        \\ 
N$_2$D$^+$       &          &  13$^a$       & 5.4 10$^{11}$ &    &
N$_2$D$^+$       &          & 22$^a$       & $<$ 2.3 10$^{11}$  &    \\
DCO$^+$           &           &  8.5             & 1.8 10$^{12}$  &  &
DCO$^+$           &         & 17           &  4.2 10$^{11}$  &   \\
D$_2$CO           &           &  10$^a$        & 2.0 10$^{12}$ &  &
D$_2$CO          &        & 11$^a$   & 3.7 10$^{12}$  &       \\  
\hline
\end{tabular}

\noindent
$^a$ Assumed rotation temperature.

\noindent
$^p$ Assuming a point source.

\noindent
$^e$ Assuming a beam filling factor of 1.
\end{table*}

\section{Analysis}
The data have been analysed using the rotation diagram method. The molecular
constants, the upper state energies and the partition functions required for applying
this method have been taken from the JPL line catalog \citep{pic98}. 
This method gives the rotation 
temperature and total column density of a particular species knowing the integrated line 
intensities of several lines with different upper state energies. The rotation temperatures
and column densitites estimated in this way are shown in Table~4. The rotation temperature
is a lower limit to the gas kinetic temperature, and only constitutes a
good measure of the kinetic temperature if the lines are thermalized. This is
the case of the ammonia inversion lines which are thermalized 
with densities, n$\sim$10$^3$~cm$^{-3}$. For this reason, we have used the
NH$_3$ inversion lines to estimate the gas kinetic temperature of the different 
envelope components.

For some molecules we have only observed one transition. In this case we have
calculated the total column density assuming optically thin emission, local thermodynamic
equilibrium (LTE) and the rotation temperature derived from a molecule with similar
excitation requirements. In these cases we have marked with the superindex $``$a"  the
rotation temperature in Table 4.

The molecules NH$_3$, N$_2$H$^+$, CN and HCN present hyperfine splitting. This allows us
to derive the line opacity directly from the hyperfine line ratios. In these cases the
column densities have been estimated directly from the line opacities. 
For N$_2$H$^+$ and NH$_3$, we have also calculated the source 
size assuming the rotation temperature derived from the NH$_3$ inversion lines when
the opacity of the main component is determined.
We have made LVG calculation for SiO, CS and C$^{34}$S. These calculations are
shown in Tables 6 and 8. 
 In all the
cases, we have fitted the densities assuming a fixed kinetic temperature. 
The assumed gas kinetic temperatures are based 
on those derived from the  NH$_3$ data in Section 5.1 and 
are shown in Table 6 and 8.  We have used the CS collisional coefficients 
calculated by \citet{gre78} in the LVG calculations. The same collisional coefficients
are used for SiO. This is a reasonable approximation since both molecules have
the same mass and similar dipole moments. 
In the case of CS, we have been able to calculate the
opacity and the source size because we have observed the main isotope and the rarer isotope
C$^{34}$S. All the column densities in Table 4, 6 and 8 are beam averaged
column densities.

\section{Results}
  
\subsection{NH$_3$} 
We have observed the ammonia  (1,1), (2,2), (3,3) and (4,4) inversion transitions
towards NGC~7129~--~FIRS~2 and LkH$\alpha$~234. All the spectra 
are shown in Fig. 1. The (1,1), (2,2) and (3,3) lines have been detected in both sources but
only an upper limit has been obtained for the (4,4) line. 
We have fitted these lines using the NH$_3$ procedure in the CLASS
(this is the program dedicated to the analysis of single-dish spectra of the GILDAS softwares
(http://www.iram.fr/IRAMFR/GILDAS). 
This procedure fit all the hyperfine
components assuming equal excitation temperature, central velocity and linewidth. 
The parameters given by
the procedure are $(T_{ex}-T_{bg})$$\times$$\tau_m$, $V$, $\Delta$V, 
and $\tau_m$ where $\tau_m$ is
the main  group opacity and $T_{ex}$ the excitation temperature (see Bachiller et al. 1987 for a 
more detailed description of this procedure). 
In case of optically thin emission $\tau_m$ cannot be determined and is set
arbitrarily to 0.1. 

The fits to the NH$_3$ lines in NGC~7129--FIRS~2 are shown in Table 2.
The NH$_3$ emission is optically thin in all the lines. However, the linewidth of
the (3,3) line is almost  a factor of 3 larger than the linewidth of the (1,1) and (2,2) 
lines (see Table 2 and Fig. 2). In addition, the central velocity of the lines is slightly lower than those of
the lower energy lines. This suggests that the (3,3) line is arising in a different 
region that the (1,1) and (2,2) lines. This interpretation is reinforced by 
the NH$_3$ rotational diagram (see Fig. 3).
The three lines detected in NGC~7129--FIRS~2 cannot be fitted by one single
straight line which implies the existence of at least two gas components with
different rotation temperatures, a cold component traced by the (1,1) 
and (2,2) lines and a hot component only detected with the (3,3)
line. Using the (1,1) and (2,2) lines, we derive a rotation 
temperature, T$_{12}$=13 K and a column density, 
N(NH$_3$)=4.9~10$^{14}$~cm$^{-2}$ for the cold component. 
Because of the lack of radiative transitions
between different K-ladders, the ammonia inversion lines are good thermometers of
dense clouds. In fact, detailed radiative transfer
calculations for NH$_3$ show that T$_k$=T$_{12}$  for T$_k$$<$20~K
\citep{dan88}. This low value of the kinetic temperature in the cold
envelope implies that depletion could  be important in this young protostar. 

Since we have not detected the (4,4) line, we
can only derive lower and upper limits for the rotation temperature of
the hot component. The lower limit is given for the excitation temperature between
the (1,1) and (3,3) lines, and the upper limit is given for the excitation temperature
between the (3,3) line and the upper limit to the (4,4) line. 
We obtain that the hot component has a rotation temperature,
T$_{rot}$$\sim$30 --90 K, which implies a lower limit of 50 K to the kinetic 
temperature of this component.  We have estimated N(NH$_3$)$\sim$1.5 10$^{13}$ cm$^{-3}$
for the hot component. This value has been derived from the integrated intensity of the (3,3) line
assuming T$_{rot}$=90 K and LTE conditions and is, in fact, a lower limit
to the actual column density of the hot gas.
  
The same procedure has been repeated for the ammonia lines towards
LkH$\alpha$~234. Similarly to NGC~7129~--~FIRS~2,  the ammonia emission towards 
LkH$\alpha$~234 cannot be fitted with one single rotation temperature
(see Fig. 3).
However, in this case the linewidth of the (3,3) line is more similar to
those of the (1,1) and (2,2) lines (see Table 3). The linewidths seem to
increase monotonically with the energy of the transition, but there is not a jump
between the linewidth of the (3,3) line and those of the others as in the
case of the protostar. 
We have fitted a two components model to the rotational diagram in 
LkH$\alpha$ 234 and
obtained rotation temperatures of  $\sim$ 22 K and 
$\sim$49--134 K for cold and hot components respectively. 
This implies a kinetic temperature of T$_k$$\sim$28 K for the cold envelope
\citep{dan88} and a lower limit of T$_k$$>$100~K for the hot component.
Contrary to  FIRS~2, the (1,1) line is moderately thick 
in this source. This allows us to estimate
the size of the NH$_3$ emission. Assuming that the excitation
temperature of the (1,1) line is equal to the rotation temperature
of the cold component, we obtain 
a size of $\approx$ 8$''$, which is lower than the size of the
clump in the 1.3mm continuum emission.  
				
\setlength\unitlength{1cm}
\begin{figure}
\vspace{10cm}
\includegraphics{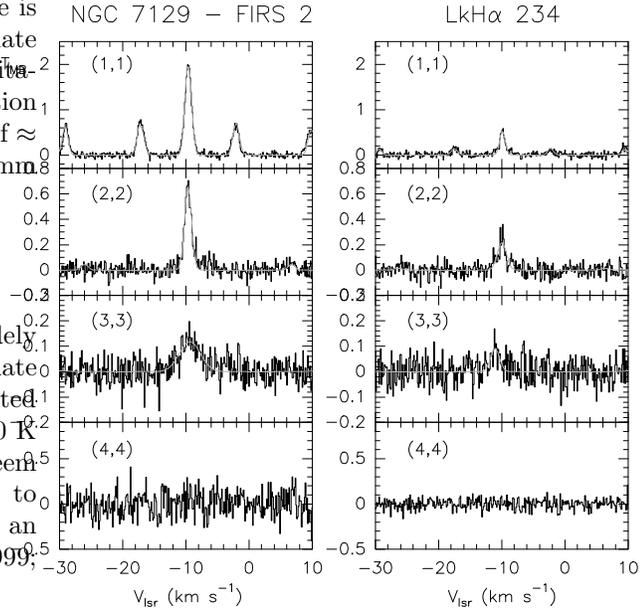}
%\end{picture}
\caption{Spectra of the NH$_3$ (1,1), (2,2), (3,3) and (4,4) inversion lines towards
 NGC~7129~--~FIRS~2 (left) and LkH$\alpha$~234 (right).}
\end{figure}
		
\setlength\unitlength{1cm}
\begin{figure*}
\vspace{7cm}
\includegraphics{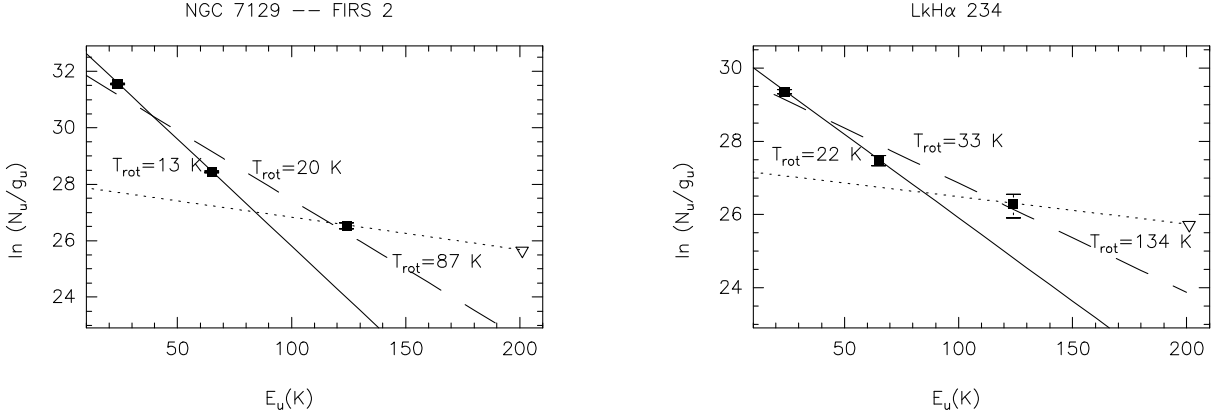}
%\end{picture}
\caption{Rotational diagram of NH$_3$ in NGC 7129 -- FIRS 2 and LkH$\alpha$ 234. The long-dashed line corresponds to
the fit assuming a single rotation temperature for all the transitions. The solid and dotted lines are the warm and cold
component respectively of the two-components model.}
\end{figure*}
		
\subsection{N$_2$H$^+$, H$^{13}$CO$^+$}

Recent studies in pre-stellar cores reveal that the widely used tracers of dense gas, C$^{18}$O
and CS, are not adequate to trace dense cold clumps. Both are strongly depleted in the core 
at densities of a few 10$^4$ cm$^{-3}$ and T$_k$$<$20 K \citep{taf04} . Only nitrogenated molecules 
seem to be unaffected by depletion. In fact, N$_2$H$^+$ seems to keep a constant abundance through the core 
and is an excellent tracer of the cold gas \citep[e.g.][]{cas99,taf02}. 

We have made small maps in the N$_2$H$^+$ 1-0 and the H$^{13}$CO$^+$ 1--0 lines 
toward FIRS 2 and LkH$\alpha$ 234. In addition we have
observed the HC$^{18}$O$^+$ line towards the star position to have an estimate of the
opacity of the H$^{13}$CO$^+$ 1--0 line. In Fig. 4 and 5 we show the  N$_2$H$^+$ 1-0 and 
the H$^{13}$CO$^+$ 1--0 line integrated intensity maps together with the continuum map at 1.3mm,
and in Table 2 and 3 we show the fits to the molecular lines. Due to the splitting of
the N$_2$H$^+$ 1$\rightarrow$0 line, we can have an estimate of the line opacity using the
same method as in the case of NH$_3$.

Intense centrally-peaked emission is observed in the continuum map at 1.3mm toward FIRS~2.  
The same morphology is observed in the H$^{13}$CO$^+$ and N$_2$H$^+$ maps, although 
the profile of the H$^{13}$CO$^+$ emission is flatter than that of N$_2$H$^+$. To quantify
this difference in the emission profile,
we have calculated the line intensity ratio, $r_1$= $I_m(N_2H^+)/I(H^{13}CO^+)$, 
where $I_m(N_2H^+)$ is the intensity of the main component of the N$_2$H$^+$ 1$\rightarrow$0 line
and $I(H^{13}CO^+)$ the intensity of the  H$^{13}$CO$^+$ 1$\rightarrow$0 line,
in a radial strip at 0$''$ offset in declination. The ratio r$_1$ changes from 2.7 at the 
center of the clump to 1.5 at an offset of ($-$30$''$,0$''$). Using the LTE approximation 
with  T$_{rot}$ = 13 K (based on our NH$_3$ calculations) and
assuming optically thin emission in the H$^{13}$CO$^+$ line, we derive that 
the N$_2$H$^+$/H$^{13}$CO$^+$ abundance ratio changes from $\sim$ 17 in the 
(0,0) position to $\sim$ 5 at (-30$''$,0 ). Thus, the abundance of H$^{13}$CO$^+$ relative
to N$_2$H$^+$ seems to decrease by a factor of $\sim$ 3 towards the clump center.
However, this change in the estimated H$^{13}$CO$^+$ abundance could
be due to the larger opacity of the H$^{13}$CO$^+$ line. To constrain the opacity
of the  H$^{13}$CO$^+$ 1$\rightarrow$0 line, we have observed the 
HC$^{18}$O$^+$ 1$\rightarrow$0 line towards the (0,0) position. 
The H$^{13}$CO$^+$ 1$\rightarrow$0 /HC$^{18}$O$^+$ 1$\rightarrow$0 
line intensity ratio is $\sim$ 11, showing that the H$^{13}$CO$^+$ 1--0 line is optically thin
at this position.

In the above calculations, we have assumed that the rotation temperature is uniform in all the strip. This
assumption could be unrealistic since the density and kinetic temperature 
are expected to decrease with
the distance from the star.
We have used an LVG code to investigate if the different values of $r_1$ are
due to a gradient in the physical conditions in the clump or/and 
are the result of a change in the relative abundance of both molecules. 
Since we are comparing the ground state lines of both species, 
the ratio $r_1$ is very little dependent
on the kinetic temperature and depends mainly on the hydrogen density and 
the N$_2$H$^+$/H$^{13}$CO$^+$ abundance ratio. In Fig. 6
we show the ratio $r_1$ and the opacity of the main hyperfine N$_2$H$^+$ 1$\rightarrow$0 line
for a wide range of physical conditions assuming   
X= N$_2$H$^+$/H$^{13}$CO$^+$=3,7,15. 
The values of the opacity and $r_1$ measured towards the star position can only be
fitted assuming X=15 (see Fig. 6). A lower value of X would imply that the N$_2$H$^+$ line
is optically thin in contradiction with our observational results. On the contrary, the value measured at the
offset (--30$''$,0$''$) can only be fitted with X$<$ 7. Thus we conclude that the gradient in the value
of $r_1$ cannot be due to the expected gradient in the excitation temperature of the observed
lines across the clump, but
to a gradient in the N$_2$H$^+$/H$^{13}$CO$^+$ abundance ratio. 

As commented above,
detailed studies in pre-stellar clumps show that the abundance of N$_2$H$^+$ remains
constant in these cold clumps while H$^{13}$CO$^+$ could suffer from depletion in
the densest part \citep{lee03,cas02}. Assuming that N$_2$H$^+$ has a constant
abundance in the clump, we need to assume an H$^{13}$CO$^+$ depletion factor $\gsim$2
to fit our observations.

The integrated intensity maps of the H$^{13}$CO$^+$ 1$\rightarrow$0 and N$_2$H$^+$ 1$\rightarrow$0 lines 
towards LkH$\alpha$ 234 are shown in Fig. 5. The N$_2$H$^+$
1$\rightarrow$0 map presents a very different morphology to that of the continuum and the H$^{13}$CO$^+$ 
maps. While the continuum map presents a intense point-like emission at the position of LkH$\alpha$ 234,
the N$_2$H$^+$ map peaks at the offset (--6$''$,+18$''$). The map of the H$^{13}$CO$^+$ line show
both peaks. As we will argue in the following, these different morphologies can only be explained if one
assumes a different temperature and chemistry between both peaks. 

In Table 4 we show the estimated N$_2$H$^+$ and H$^{13}$CO$^+$ column densities
in LKH$\alpha$~234 assuming a rotation temperature of 22 K derived
from the NH$_3$ observations. We have obtained N(N$_2$H$^+$)$\sim$1.0 10$^{13}$ cm$^{-2}$
and a N$_2$H$^+$/H$^{13}$CO$^+$ abundance ratio of $\sim$6. The N$_2$H$^+$/H$^{13}$CO$^+$
ratio in LkH$\alpha$ 234 is lower  than that found in FIRS~2. 

We have repeated the same calculations for the offset (-6$''$,+18$''$). Continuum observations
suggests that this clump is colder than that associated to LkH$\alpha$~234. In fact, if
we assume that the N$_2$H$^+$ abundance is uniform in the region and a typical dust
temperature of 30~K towards  LkH$\alpha$~234, we need to assume a dust temperature
$\lsim$10~K at the offset  (-6$''$,+18$''$) in order to explain the measured continuum flux.
Thus, we have assumed a rotation lower temperature, T$_{rot}$$\sim$10~K,
in our LTE caculations. With these assumptions, we obtain a N$_2$H$^+$ column density 
of $\sim$1.3~10$^{13}$~cm$^{-2}$ and a N$_2$H$^+$/H$^{13}$CO$^+$ abundance ratio of $\sim$10.
This value is intermediate between those measured in LkH$\alpha$~234 and FIRS~2.
However, the difference is of a factor of 2 which is within the uncertainty involved in these kind of 
calculations. Based on our results, we can propose an evolutionary trend based on the N$_2$H$^+$/H$^{13}$CO$^+$ ratio.
This ratio is maximum in the IM Class 0 object FIRS 2 where molecular depletion
is significant, it may take an intermediate value in the infrared low-mass star IRS~6 and
is minimum in the envelope of the more evolved object, the HBe star LkH$\alpha$~234.

\begin{table*}[]
\caption{Beam averaged N$_2$H$^+$ and H$^{13}$CO$^+$ column densities}
\begin{tabular}{l cccc}
\\ \hline
\multicolumn{1}{c}{Source} & 
\multicolumn{1}{l}{T$_{rot}$ (K)} &       
\multicolumn{1}{c}{N(H$^{13}$CO$^+$)} &
\multicolumn{1}{c}{N(N$_2$H$^+$)} &
\multicolumn{1}{c}{N$_2$H$^+$/H$^{13}$CO$^+$}    \\
\multicolumn{1}{c}{} & 
\multicolumn{1}{l}{(K)} &       
\multicolumn{1}{c}{(cm$^{-2}$)} &
\multicolumn{1}{c}{(cm$^{-2}$)} &
\multicolumn{1}{c}{}   \\
\hline
FIRS 2 & 13 & 2.2 10$^{12}$ & 3.8 10$^{13}$ &  17  \\
(--30,0)                   & 13 & 8.0 10$^{11}$ & 4.2 10$^{12}$ &  5    \\
LkH$\alpha$ 234     &  22 & 1.7 10$^{12}$  & 1.0 10$^{13}$  &  6   \\
(--6,+18)                 &  10 & 8.8 10$^{11}$  & 8.5 10$^{12}$  &   $\sim$ 10     \\    \hline
\end{tabular}
\end{table*}

\setlength\unitlength{1cm}
\begin{figure*}
\vspace{17cm}
\includegraphics{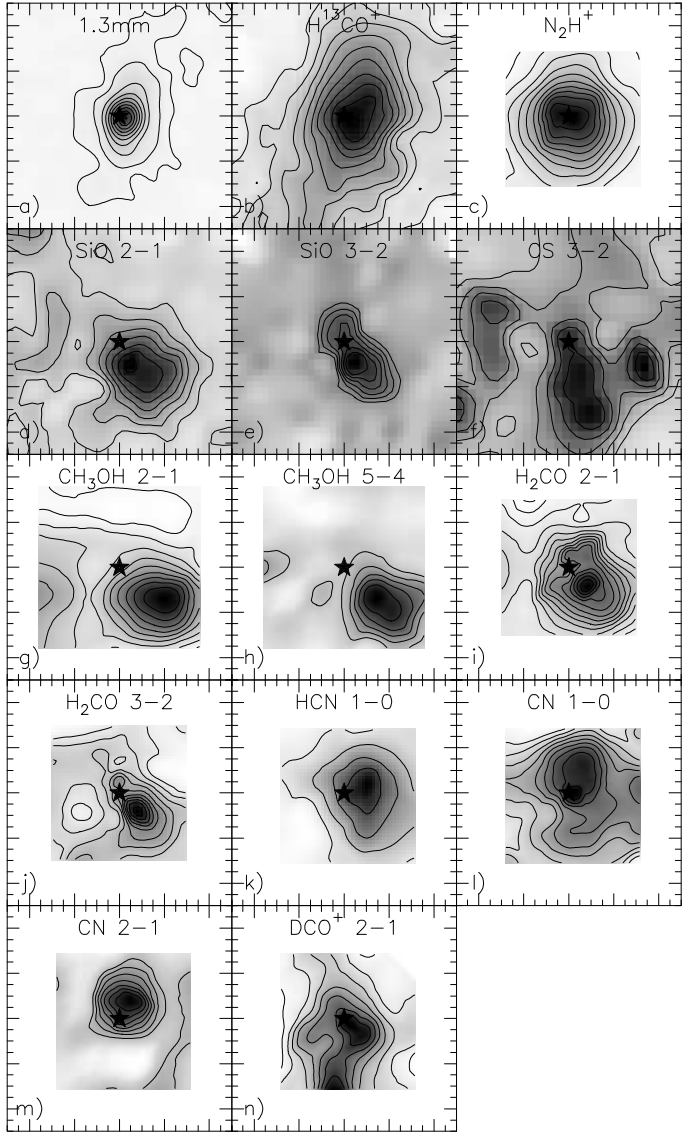}
%\end{picture}
\caption{Continuum map at 1.3mm and integrated intensity maps of the 
observed lines in NGC~7129~--~FIRS~2. The box size is 100$''$$\times$100$''$
and is centered at the star position.
Contour levels are: a) 39.2 mJy/beam, and from 78.5 to 706.2~mJy/beam by  
78.5~mJy /beam ;b) 0.19 to 1.7 by  0.19~K~kms$^{-1}$;
c) 0.2 to 8.2 by 0.8~K~kms$^{-1}$. d) 0.25 to 4 by 0.5~K~kms$^{-1}$;
e) 1.5 to 4 by 0.5~K~kms$^{-1}$; f) 1.5 to 13.5 by 1.0~K~kms$^{-1}$;
g) 2 to 20 by 2~K~kms$^{-1}$; h) 0.1 to 0.5 by 0.1~K~kms$^{-1}$;
i) 1 to 14 by 1~K~kms$^{-1}$; j) 1 to 21 by 2~K~kms$^{-1}$;
k)  2 to 7 by 1~K~kms$^{-1}$; l)  0.2 to 1.9 by 0.2~K~kms$^{-1}$;
m)  0.8 to 3.4 by 0.4~K~kms$^{-1}$; n)  0.2 to 3 by 0.4~K~kms$^{-1}$.}
\end{figure*}
		
\setlength\unitlength{1cm}
\begin{figure*}
\vspace{11cm}
\includegraphics{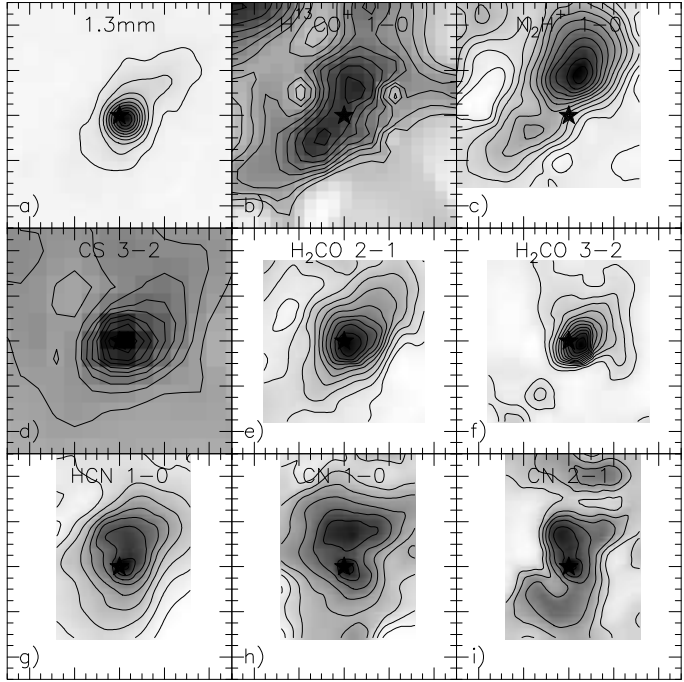}
%\end{picture}
\caption{Same as Fig. 1 for LkH$\alpha$ 234. Contour levels are:
a) 39.2 mJy/beam, 78.5 to 706.2 mJy/beam by steps of 
78.5 mJy /beam ; b) 0.19 to 1.7  by 0.19 K kms$^{-1}$;
c) 0.2 to 8.2 by 0.4 K kms$^{-1}$; d) 1.5 to 15 by 1.5 K kms$^{-1}$;
e) 1 to 11 by 1 K kms$^{-1}$; f) 1.5 to 16.5 by 1.5 K kms$^{-1}$;
g) 2 to 11 by 1 K kms$^{-1}$; h) 0.5 to 5 by 0.5 K kms$^{-1}$;
i) 1 to 6 by 1 K kms$^{-1}$}
\end{figure*}
		
\setlength\unitlength{1cm}
\begin{figure*}
\vspace{6cm}
\includegraphics{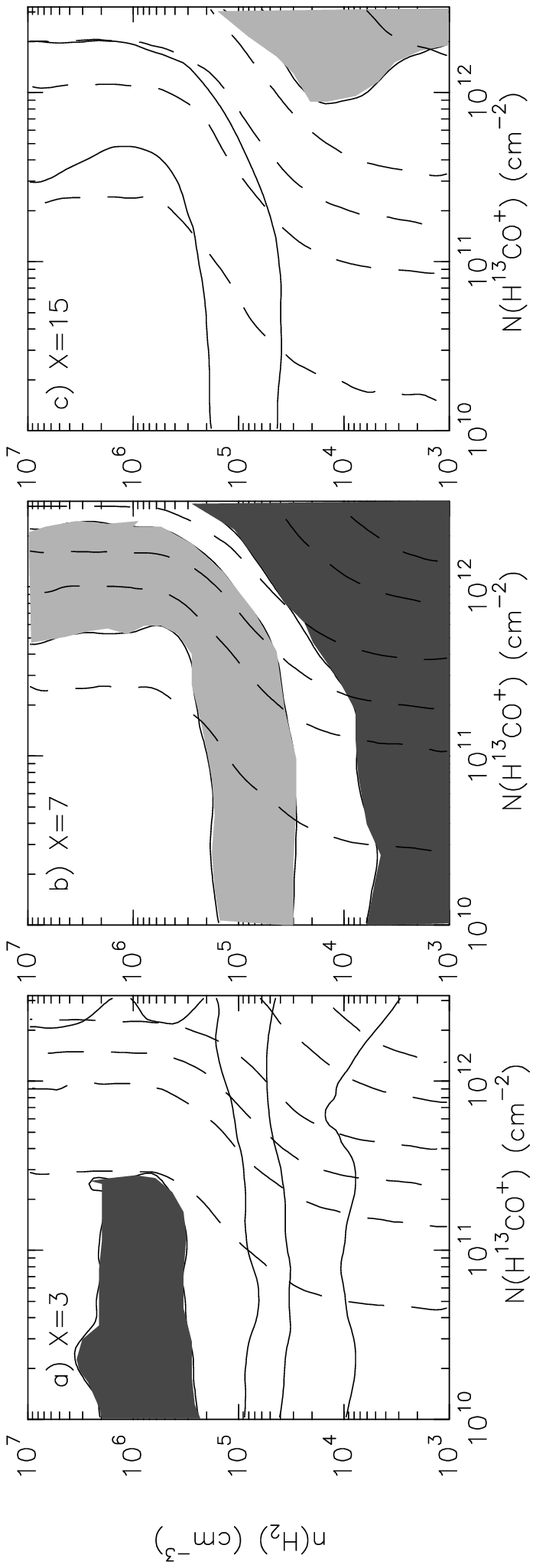}
%\end{picture}
\caption{Plots of the $r_1=I(N_2H^+)/I(H^{13}CO^+)$ ratio and opacity of the 
main N$_2$H$^+$ hyperfine component (solid contours),
and $\tau_m$ (dashed contours), as a function of the H$^{13}$CO+ column density, 
$N(H^{13}CO^+)$, and hydrogen density, n, for a kinetic 
temperature of $T_k=15 K$ and an N$_2$H$^+$/H$^{13}$CO$^+$ abundance ratio, X=3,7,15. We have shaded with dark grey
the region with $r_1$ = 1.5$\pm$0.3 ( ratio measured at the 32$''$ offset from the star position) and with a clear color
$r_1$ = 2.7$\pm$0.6 (ratio measured at the star position). We need a depletion of at least a factor of 2 towards the
center in order to fit both $tau_m$ and $r_1$ in both positions. Contour levels are: a) $r_1$ = 0.8, 1, 1.2 and
$\tau_m$ = 0.1, 0.5, 1.0, 2, 5,10; b) $r_1$ = 1.8, 2.2, 2.7, 3.2 and $\tau_m$ = 0.1, 0.5, 1.0, 2, 5, 10; 
c) $r_1$ = 3.2, 5, 7 and
$\tau_m$ = 0.1, 0.5, 1.0, 2, 5,10 }
\end{figure*}

\subsection{CO outflows, SiO}
The SiO abundance is strongly enhanced (up to several orders of magnitude) in shocked
regions. However, its abundance is very low in dark clouds and PDRs \citep{sch01}. 
Because of this peculiarity, SiO is used as a diagnostic for shocks in both galactic and
extragalactic regions \citep[e.g.][]{bac91,mar92,gar00,gar01}. 
In particular, it is commonly used to look for the energetic 
outflows associated with the youngest stellar objects. 
We have carried out  maps of the  J=2$\rightarrow$1 and
3$\rightarrow$2 rotational lines of SiO around FIRS 2 and LkH$\alpha$ 234  
to study the physical conditions and chemistry of the bipolar outflows found in these regions.

Interferometric and single-dish CO 2$\rightarrow$1 observations show the existence of two
bipolar outflows associated with FIRS 2. 
The axis of these outflows form an
angle of almost 90$^\circ$ giving a quadrupolar morphology to the spatial
distribution of the high velocity gas of the region. In Fig. 7 we show the maps of the 
high velocity $^{12}$CO 2$\rightarrow$1 emission reported by \citet{fue01}.
Interferometric observations shows
that the NE-SW outflow, hereafter FIRS 2-out 1, is the one associated with the millimeter source
FIRS 2-MM1. 
An intense blue lobe is detected in the CO 2$\rightarrow$1 emission in this outflow while 
the red lobe is only marginally detected. The NW-SE outflow, hereafter FIRS 2-out 2, seems to
be associated with an infrared star FIRS 2-IR \citep{fue01}.  In FIRS 2-out 2, the red lobe is more prominent 
than the blue one, giving a peculiar appearance to the high velocity gas around FIRS~2.
The peaks of the $^{12}$CO emission in this lobe present a $``$Z"-like shape which suggests
that the axis of the outflow FIRS 2-out 2 is
precessing. Precession has been found in bipolar outflows associated to young
low-mass stars \citep{gue96}. 

In Fig.~8 we present the integrated intensity map of the SiO 2$\rightarrow$1 line towards
FIRS~2. The SiO emission is only observed along the axis of the two outflows. 
Furthermore the SiO peaks are not located at the star
position but in the bow shocks located at the end of the outflow lobes.
Thus, the SiO emission seems to be closely related to the bipolar outflows. The spectra of
the SiO lines towards some selected positions are shown in Fig.~8.
The SiO  2$\rightarrow$1 line towards the star position presents two velocity components, a
narrow component with $\Delta v$ $\sim$ 1.6 km s$^{-1}$ which is centered at
the velocity of the ambient cloud, $V_{lsr}$$\sim$ $-10.0$km s$^{-1}$, and a
wide component,  $\Delta v$$\sim$7.0 km s$^{-1}$  which is centered at  
$V_{lsr}$$\sim$$-7.0$$\pm$$1.0$ km s$^{-1}$. The component at 
 $-7.0$ kms$^{-1}$ corresponds to a well defined high velocity clump
which peaks at the position (-10$''$,-10$''$) (see Fig.~8). Hereafter we will
refer to this clump as R1. There exists a counterpart blue clump  
at a velocity, V$_{lsr}$$\sim$ $-12.5$$\pm$$0.5$ km s$^{-1}$ which 
peaks at the position (50$''$,50$''$) (hereafter B1). At this position, the 
profile of  the SiO 2$\rightarrow$1 line only presents a wide component
with $\Delta v$$\sim$7~km~s$^{-1}$. The high velocity
clumps R1 and B1 have well defined velocities and positions like the $``$bullets"
found in low-mass stars \citep[see][]{bac96}.  The jet-like morphology of the SiO emission along
the outflow axis as well as the existence of bullets argues in favor of the 
youth of this outflow.

The component at the velocity of the ambient cloud, $V_{lsr}$$\sim$$-10.0$ km s$^{-1}$,
is also present in FIRS 2-out 2. The emission of this narrow component in FIRS 2-out 2 is 
surrounding the red lobe as detected in $^{12}$CO. In fact, the narrow component is 
located adjacent to the peaks of the high velocity CO emission suggesting that SiO 
emission is tracing the molecular cloud gas entrained in the outflow. Narrow SiO components 
widespread in the molecular cloud in which the bipolar outflow is embedded
has been detected in other Class 0 protostars \citep{cod99}.  
We propose an interpretation in which the
morphology of the SiO emission is related to the evolutionary stage of the outflow. In FIRS~2-out~1, 
the SiO emission has a jet-like morphology and is concentrated in $``$bullets" ejected by the exciting 
star. This $``$jet-like" morphology is also observed in the interferometric $^{12}$CO image reported by
\citep{fue01}. The SiO emission in FIRS~2-out~2 is surrounding the red CO lobe. We propose that
in this case the SiO emission traces the material adjacent to the cavity walls excavated by the
outflow which is being entrained into the outflow. The different profiles of the SiO emission 
in FIRS~2-out~1 and FIRS~2-out~2 are clearly seen in Fig.~8.  

We have calculated the  SiO 3$\rightarrow$2/2$\rightarrow$1 line intensity ratio,
$r_{32}$, at the (0$''$,0$''$) position by degrading
the angular resolution of the SiO 3$\rightarrow$2 map to that of the SiO 2$\rightarrow$1 one.  
A value of $r_{32}$$\sim$ 0.8 is found in the three channels centered
at the ambient velocity, while a value $r_{32}$ $>$ 1.0 is found for the high
velocity gas. This reveals a higher excitation temperature for the high velocity gas. In order
to determine the physical conditions for both components, we have used an LVG 
code. Since $r_{32}$ suggests different
physical conditions for the different velocity ranges, we have carried out LVG calculations for the ambient,  
moderate velocity and high velocity gas components (see Table 6).
Taking into account the kinetic temperature derived from the NH$_3$ lines
we have assumed $T_k$=15 K for the ambient narrow component and T$_k$=50 K for the
high velocity components. The derived densities and column densities are shown in 
Table 6. We are aware that the NH$_3$ and SiO emissions could 
arise in different regions and the assumed kinetic temperatures are very
uncertain. But since the SiO lines are optically thin, the derived column densities
are  weakly dependent on the assumed kinetic temperature and are accurate
within a factor of $\sim$2. The values of Table~6 clearly show that the SiO abundance
is larger by almost 2 orders of magnitude in the high velocity gas of FIRS~2-out~1 than in
the narrow component associated with FIRS~2-out~2. Within FIRS~2-out~1, we also detect
a gradient in the SiO abundance, being the SiO/$^{13}$CO ratio 2 orders of magnitude
larger in the high velocity component than in the ambient component. This enhancement
of the SiO abundance in the high velocity gas is also observed in very young
low-mass protostellar outflows \citep{bac91,bac01} and interpreted as due to the chemical
gas processing by the energetic shocks associated with
the high velocity $``$bullets".

Hydrogen densities are also quite independent on the assumed
temperature for T$_k$$>$50~K. The estimated hydrogen density decreases by only a factor 
of $\sim$4 if we change the kinetic temperature from 15~K to 100~K. Thus, the estimated hydrogen 
densities are accurate within a factor of $\sim$4. Our calculations show that the density seems 
to increase from a few 10$^5$~cm$^{-3}$ in the ambient component to $>$10$^6$~cm$^{-3}$ in the high 
velocity gas at the star position. We have also carried out LVG calculations for R1 and B1, and at the
offset (50$''$,-50$''$) where only the narrow component has been detected.
The density in the bullets R1 and B1 is $\ga$10$^5$~cm$^{-3}$ while the density in the narrow
component detected at the offset (50$''$,-50$''$) is of a few 10$^4$~cm$^{-3}$. This density
is an upper limit to the hydrogen density in the narrow component, since the
kinetic temperature is never expected to be lower than 15~K. Thus, we consider that
the estimated difference in the hydrogen density of the wide and narrow components is
reliable. The lower density of the narrow component also supports our interpretation of the SiO emission 
arising in the gas of the molecular cloud surrounding the outflow.

We have observed a map of 80$''$$\times$80$''$ around LkH$\alpha$ 234
in the SiO 2$\rightarrow$1 and 3$\rightarrow$2 lines. 
We have detected SiO emission only towards the star position.
Since we have integrated double time towards the star position than
in the other map positions, we cannot exclude the possibility of SiO emission at the
same level in other positions. Thus we have
poor information about the spatial distribution of the emission. Regarding
the kinematical information, the large linewidth of the SiO
lines, $\Delta v$$>$3 km s$^{-1}$, compared to that of the (1,1) and (2,2) ammonia lines
suggests that the emission arises in the warm component.
The weakness of the SiO 2$\rightarrow$1 line emission as well as the lack of information 
about the source size make any estimate of the
hydrogen density very uncertain. We have derived the SiO column density assuming
n$\sim$ 5 10$^5$ cm$^{-3}$. With this assumption we derive a SiO column 
density of $\sim$10$^{11}$~cm$^{-2}$. 
This value is lower by a factor of $>$6 than the total SiO column
density towards FIRS 2. The SiO/$^{13}$CO ratio towards this
star is similar to that found in PDRs \citep{sch01}.

In Table 7 we present a summary of the SiO observations. There is a clear 
evolutionary trend in the SiO behavior. The youngest outflow FIRS~2-out~1
presents intense SiO emission at high velocity with SiO abundances as high
as $\sim$10$^{-8}$. Towards the more evolved outflow FIRS~2-out~2 we have
detected only a weak SiO line at ambient velocity. The SiO abundance in this
component is $\sim$10$^{-10}$, i.e., two orders of magnitude lower than the 
SiO abundance in the high velocity gas associated with FIRS~2-out~1 but still
larger than the SiO abundance in PDRs and dark clouds. 
Towards LkH$\alpha$ 234 we have detected SiO emission at the ambient velocities with a
fractional abundance of 10$^{-12}$. This abundance is similar to that measured in PDRs and
could be associated with the PDR produced by LkH$\alpha$~234.

This evolutionary trend confirms that SiO is a good tracer of energetic shocks. The SiO
abundance is highly enhanced when the shocks are strong enough to release the silicon
from the grains to the gas phase \citep{mar92}. In a MHD shock model the 
release of Si to the gas phase requires V$_{shock}$$>$ 40 km s$^{-1}$ \citep{flo96}.
This is consistent with the trend of having larger SiO abundances in the higher
velocity gas. Because of projection effects, the velocity we measure is
a lower limit to V$_{shock}$. As the protostar evolves, the ouflow fades and the amount of
high velocity molecular gas decreases. This produces a decrease in the SiO abundance.
When the bipolar molecular outflow stops, the SiO abundance around the star 
decreases to the typical value in PDRs. 

\setlength\unitlength{1cm}
\begin{figure*}
\vspace{9cm}
\includegraphics{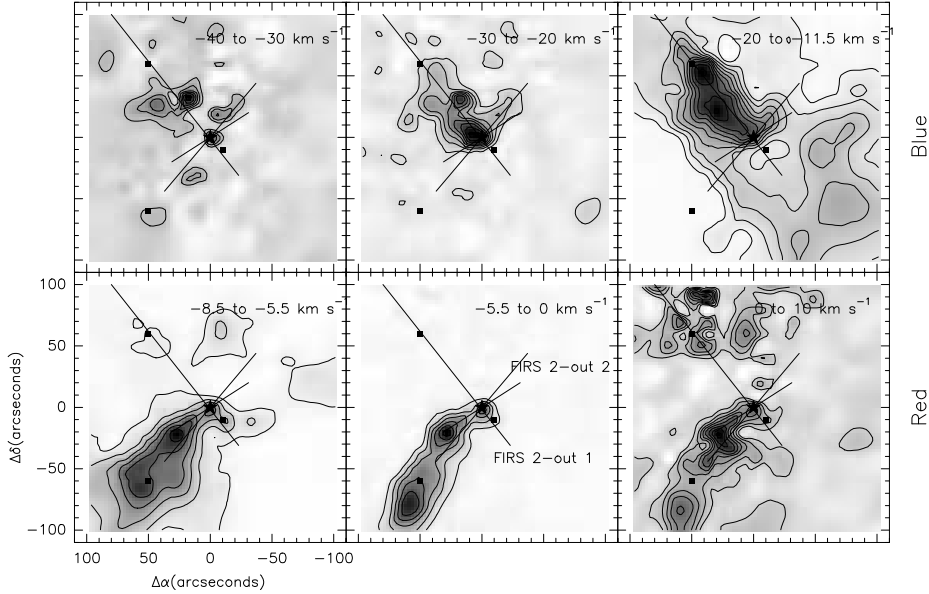}
%\end{picture}
\caption{High velocity CO emission in FIRS~2. The blue-shifted velocities
are shown in the top panels, and the red-shifted velocities in the bottom panels. Straight lines
indicate the outflow axis. In the case of FIRS~2-out~2 we have drawn two axis because of
outflow precession. Contour
levels are: a) 1 to 6 by 1~K~kms$^{-1}$; b) 2 to 22 by 2~K~kms$^{-1}$; c) lev 7 to 70 by 7~K~kms$^{-1}$; 
d) 7 to 80 by 7~K~kms$^{-1}$; e)  4 to 50 by 4~K~kms$^{-1}$;
f) 1 to 8 by 1~K~kms$^{-1}$.}
\end{figure*}
		
%\setlength\unitlength{1cm}
%\begin{figure}
%\vspace{4cm}
%\special{psfile=art_fig7.ps hoffset=-50
%voffset=150 hscale=45 vscale=45 angle=-90}
%\end{picture}
%\caption{Integrated intensity SiO maps in different velocity intervals in NGC~7129~--~FIRS~2.}
%\end{figure}
		
\setlength\unitlength{1cm}
\begin{figure}
\vspace{18cm}
\includegraphics{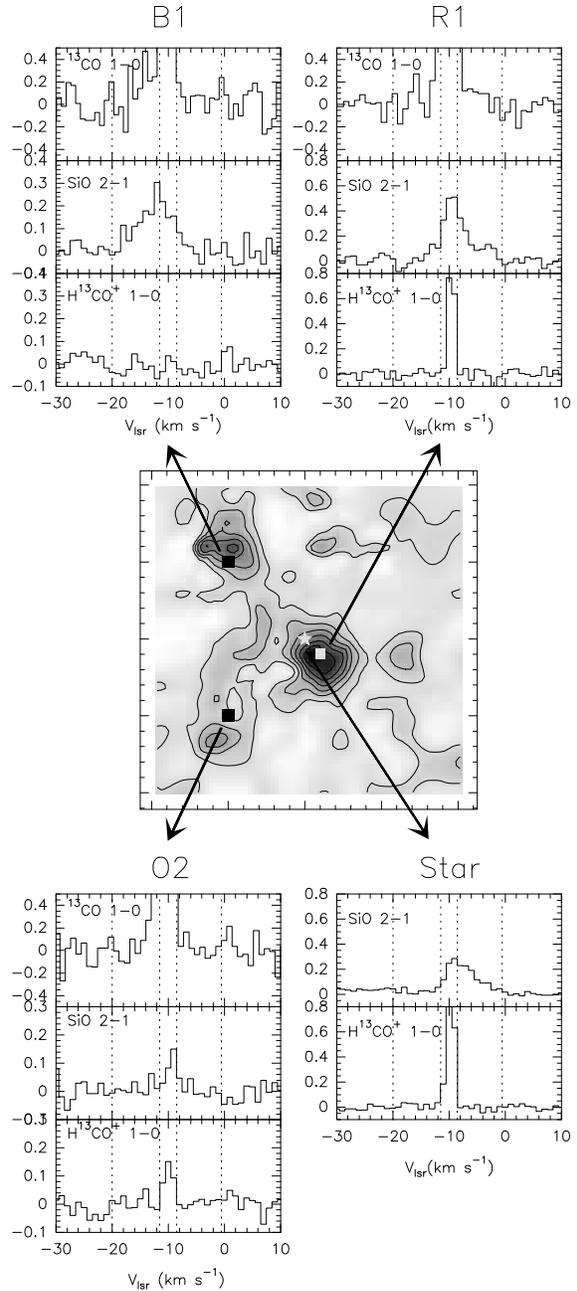}
%\end{picture}
\caption{In the central panel we show the integrated line intensity map of the SiO J=2$\rightarrow$1 line toward
FIRS~2. Contour levels are from 0.25~K~kms$^{-1}$ to 4~K~kms$^{-1}$ by steps of
0.5~K~kms$^{-1}$. Surrounding the central panel, we show the observed spectra of
the $^{13}$CO 1$\rightarrow$0, SiO 2$\rightarrow$1 and H$^{13}$CO$^+$ 1$\rightarrow$0 lines
towards selected positions of outflows FIRS~2-out~1 (top panels) and FIRS~2-out~2 (bottom panels). The intensity
scale in the spectra is main brightness temperature.}
\end{figure}

\begin{table*}[]
\caption{LVG calculations : SiO}
\begin{tabular}{ll ccccccc}
\\ \hline
\multicolumn{1}{c}{Source} & \multicolumn{1}{c}{Vel. range} &
\multicolumn{1}{c}{T$_k$(K)} & 
\multicolumn{1}{c}{n} & 
\multicolumn{1}{c}{N($^{13}$CO)} &
\multicolumn{1}{l}{N(SiO)} &
\multicolumn{1}{c}{N(H$^{13}$CO$^+$)} & \multicolumn{2}{c}{} \\ 
\multicolumn{1}{c}{}  & \multicolumn{1}{c}{(km s$^{-1}$)} &
\multicolumn{1}{c}{(K)} & 
\multicolumn{1}{c}{(cm$^{-3}$)} & \multicolumn{1}{l}{(cm$^{-2}$)} &
\multicolumn{1}{c}{(cm$^{-2}$)} &  
\multicolumn{1}{c}{(cm$^{-2}$)} &
\multicolumn{1}{c}{SiO/$^{13}$CO} &
\multicolumn{1}{c}{SiO/H$^{13}$CO$^+$} \\ 
\hline
FIRS 2 (Star)  & -13.5 -- -11.5 & 50 & 5 10$^5$      &    &  1.5 10$^{11}$  & 1.0 10$^{11}$ &  &  1.5 \\
                    & -11.5 -- -8.5   & 15    &  5 10$^5$   &   &  1.1 10$^{12}$ & 1.5 10$^{12}$ &  & 0.7    \\
                    & -8.5 -- -4.5     & 50  & 4 10$^5$      &    & 1.2 10$^{12}$ &   &  &  \\
                    & -4.5 -- -0.5     & 50  & 1 10$^6$      &    & 4.7 10$^{11}$  &   &  &  \\
                    & Total              &       &                   &     & 2.8 10$^{12}$ & 1.5 10$^{12}$  &  & 1.9  \\ \\
(-10$''$,-10$''$) R1  & -13.5 -- -11.5 &  50      &  5 10$^5$$^*$  & 3 10$^{15}$   & 4 10$^{11}$  &  & 0.0001 &    \\
 (FIRS 2-out 1)   & -11.5 -- -8.5  &  15      &   4 10$^5$        & 4 10$^{16}$   & 3 10$^{12}$  & 1 10$^{12}$ & 0.00007 & 3   \\
                       & -8.5 -- -4.5    &  50     &   4 10$^5$        & 3  10$^{15}$  & 1.5 10$^{12}$  & & 0.0005 &   \\
                        & -4.5 -- -0.5     &  50    &   3 10$^4$        & 3 10$^{14}$   & 1.5 10$^{12}$  &  & 0.005 & \\
                         & Total             &           &                      &  4.6 10$^{16}$ & 6.4 10$^{12}$ & 1 10$^{12}$ & 0.0001 & 6   \\ \\
(50$''$,50$''$) B1   &  -17.5 -- -12.5  & 50     &  2 10$^5$     & 1.5 10$^{15}$     & 1.5 10$^{12}$ & &  0.001 &        \\
(FIRS 2-out 1)         &  -12.5 -- -8.5    & 50     &  2 10$^5$     & 5.5 10$^{16}$     & 1.5 10$^{12}$ & &  0.00003  &      \\
                            & -8.5 -- -4.5       & 50     &  2 10$^5$     & 2.0 10$^{15}$     & 2.0 10$^{11}$ &  &  0.0001      \\
                              & Total          &          &                    & 5.8 10$^{16}$     & 3.2 10$^{12}$  & $<$ 2 10$^{12}$  & 0.00005 & $>$ 1 
\\ \\
(50,-50) O2  &  -11.5 -- -8.5   & 15     &  8 10$^4$     &                         & 1.0 10$^{11}$ &  3 10$^{11}$  & & 3   \\ 
(FIRS 2-out 2)         &          &                     &                       &                      &                     &  &       \\ \hline
\\             
LkH$\alpha$ 234      & -12.5 -- -8.5  & 50 & 5 10$^5$$^*$ & 1.3 10$^{17}$ & 2.0 10$^{11}$  & 1.0 10$^{12}$ & 0.000001 & 0.2 \\ 
(IRS 6-out 1)              &                      &       &                     &                 &                         &                  &                &        \\ \hline                                
\end{tabular}

\noindent
$^*$ Assumed value for the hydrogen density
\end{table*}

\begin{table}[]
\caption{SiO emission}
\begin{tabular}{l ccc}
\\ \hline
\multicolumn{1}{c}{Outflow} & \multicolumn{1}{c}{Age (Myr)} & \multicolumn{1}{c}{Component} &
\multicolumn{1}{c}{X(SiO)$^3$} \\  \hline
FIRS 2-out 1 & 0.003$^1$  & High velocity & $\sim$ 10$^{-8}$  \\
FIRS 2-out 2 & 0.005$^1$ &  Ambient         & $\sim$ 10$^{-10}$ \\
LkH$\alpha$ 234   &   $\sim$ 0.1$^2$   & Ambient  & $\sim$ 10$^{-12}$--10$^ {-11}$ \\ \hline                                 
\end{tabular}

\noindent
$^1$ Estimated age of the outflow assuming a velocity of 15 km s$^{-1}$ and the lobe sizes measured from
Fig. 1.\\
$^2$ Stellar age from \citet{fue98}\\
$^3$ Fractional abundance estimates assuming X($^{13}$CO) = 2 10$^{-6}$ and X(H$^{13}$CO$^+$)=10$^{-10}$ \\
\end{table}

\setlength\unitlength{1cm}
\begin{figure}
\vspace{9cm}
\includegraphics{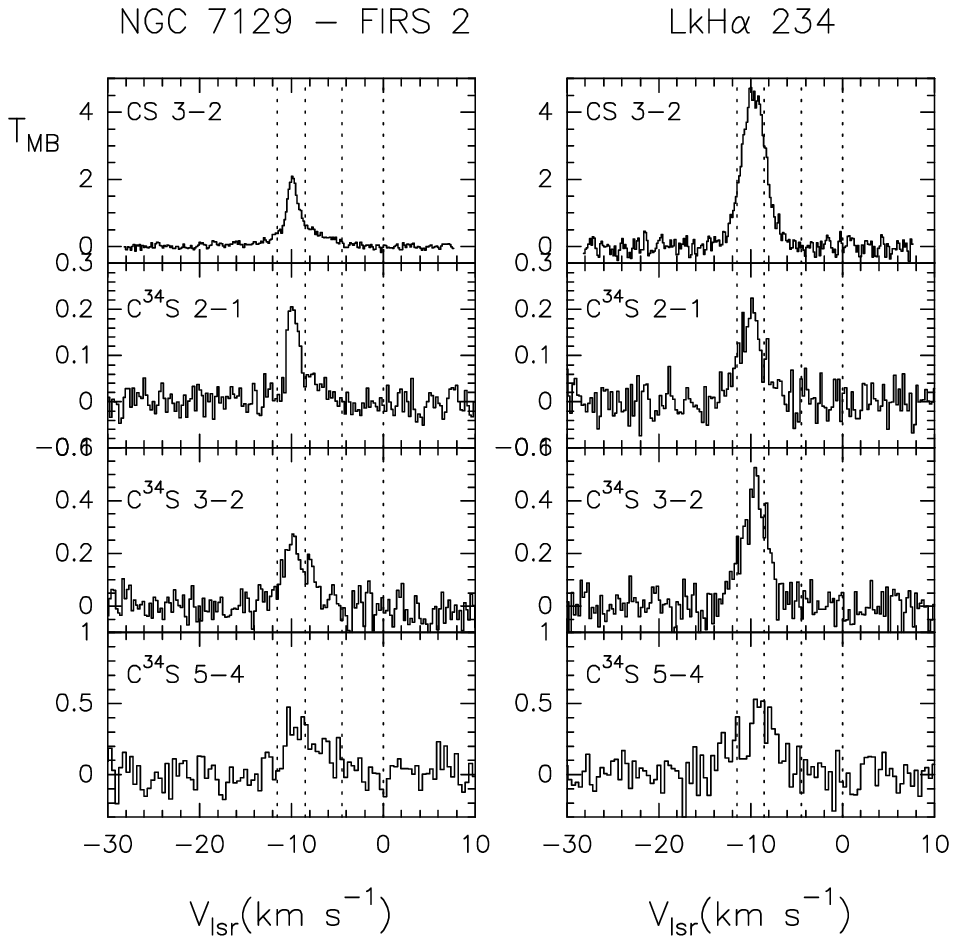}
%\end{picture}
\caption{Observed CS and C$^{34}$S spectra towards the (0,0) position in NGC~7129~--~FIRS~2 and LkH$\alpha$~234.}
\end{figure}

\subsection{CS, C$^{34}$S}
 
We have carried out a map of the CS 3$\rightarrow$2 emission towards FIRS~2 and 
LkH$\alpha$~234.
In addition we have observed the C$^{34}$S 2$\rightarrow$1, 3$\rightarrow$2 and 5$\rightarrow$4 lines towards
the (0,0) position. The spectra towards the (0,0) position are shown in Fig.~9 and the Gaussian fits
are shown in Tables 2 and 3. The integrated intensity maps of the CS J=3$\rightarrow$2 line are shown in Fig.~4.

Similarly to the case of SiO, the profiles of the CS and C$^{34}$S lines towards FIRS~2 present 
high velocity wings at red-shifted velocities. However, the terminal velocity of these wings is lower than those
of the SiO lines. The velocity range of the CS emission 
is between -14 and -5 km~s$^{-1}$, i.e. , it is only
detected at the ambient velocities and in the moderate velocity component of the SiO emission. 
The spatial distribution of this component in the CS 3$\rightarrow$2 line is different from 
that of the same component of the SiO emission,
suggesting that both molecules might be tracing different gas components even when comparing
the same velocity range.

We have carried out LVG calculations to estimate
the physical conditions of the gas emitting in CS.  
Different line ratios are found at the ambient velocities and
in the moderate velocity range in the C$^{34}$S lines (see Fig. 9). 
Thus, we have estimated the physical conditions in
the two velocity intervals. 
We have derived  a density $\ge$3~10$^6$~cm$^{-3}$ for the ambient component 
in FIRS 2. This density is larger by a factor $\sim$6 than the density 
estimated from SiO lines. This difference is not due to the assumed kinetic temperature,
since even assuming a kinetic temperature of 100 K the density would decrease 
only by a factor of $\sim$4. However, the densities derived from the C$^{34}$S lines 
in the moderate range are similar to those derived from the SiO lines. 
We propose that at least part of the C$^{34}$S lines 
emission at ambient velocities does not arise in the outflow. We speculate with the possibility of the
C$^{34}$S emission arising in a hot core.

The CS and C$^{34}$S lines observed in LkH$\alpha$~234 are shown in Fig.~9 The
line-widths of the C$^{34}$S lines are similar to that of the SiO and the 
NH$_3$ (3,3) lines suggesting that they are tracing a warm component. The emission of
the CS lines is concentrated towards the star position, therefore the connection between
 the outflow and the CS emission is not clear. We have derived a
density $\sim$10$^5$~cm$^{-3}$ for this warm component.
 This density is lower than the
typical density of the hot cores associated with massive stars. Since LkH$\alpha$ 234 is
a visible star, the envelope has already been disrupted by the star and the UV radiation
is escaping through the envelope. A dense (n$>$ 10$^6$ cm$^{-3}$) hot region 
similar to the hot cores in massive stars is not expected at this evolutionary stage.

\begin{table*}[]
\caption{LVG calculations : CS, C$^{34}$S }
\begin{tabular}{ll cccc}
\\ \hline
\multicolumn{1}{c}{Source} & \multicolumn{1}{c}{Vel. range} & \multicolumn{1}{c}{T$_k$(K)} & 
\multicolumn{1}{c}{n (cm$^{-3}$)} & \multicolumn{1}{l}{N$^s$(C$^{34}$S)(cm$^{-2}$)} 
& \multicolumn{1}{c}{$\Omega_s$} \\ \hline
NGC 7129 -- FIRS 2 & -10.5 -- -9.5 km s$^{-1}$ & 15 & 3 10$^6$  & 8.0 10$^{11}$     & $\sim$9$''$\\
                              & -9.5 -- -6.5 km s$^{-1}$   & 50 & 6 10$^5$  & 7.1 10$^{11}$    & $\sim$4$''$ \\
                              & Total                             &      &                &  1.5 10$^{12}$   &                  \\            
LkH$\alpha$ 234      & -11.5 -- -9.5 km s$^{-1}$  & 50 & 2 10$^5$ & 9.6 10$^{11}$     & 14$''$ \\
                              & -9.5 -- -5.5 km s$^{-1}$    & 50 &  4 10$^5$ & 7.8 10$^{11}$   & 6$''$ \\ 
                              & Total                              &      &                &  1.7 10$^{12}$  &       \\ \hline                                
\end{tabular}
\end{table*}

\subsection{CH$_3$OH}
We have observed up to 17 methanol lines towards the (0,0) position in
FIRS~2 and LkH$\alpha$~234. 
The large number of lines allows us to have a good estimate of the rotation temperature and 
the CH$_3$OH column density. In addition, we have obtained a small map 
in the intense CH$_3$OH~2$\rightarrow$1 and 5$\rightarrow$4 lines towards FIRS~2. 
The CH$_3$OH maps around FIRS~2 are shown in Fig.~4.
The CH$_3$OH emission peaks towards the bullet R1 
suggesting that these lines arise mainly in the molecular outflow.

The high-velocity resolution spectra of the CH$_3$OH lines are shown in Fig.~10. 
The velocity profiles of the CH$_3$OH lines towards FIRS~2 are similar to 
those found in the SiO lines. The low energy lines present two velocity components,
a narrow component at the ambient velocity and a wide wing which extends to red-shifted
velocities and is centered at -6$\pm$1 km s$^{-1}$ . In higher energy transitions,
the narrow component becomes weaker and only the wide red-shifted one is detected.
This is consistent with the integrated intensity maps of the CH$_3$OH~2$\rightarrow$1 and 
5$\rightarrow$4 lines shown in Fig.~4 which show a strong peak emission at the position of the bullet R1. 
In order to further investigate the nature of the CH$_3$OH emission,  we have fit the 
lines with Gaussian profiles and make some correlations. 
In Fig.~11 we plot the central velocity and 
linewidths of the CH$_3$OH lines vs. the upper state energy of the observed transition. 
It is clearly seen that the linewidths of the CH$_3$OH lines increase with the upper 
state energy of the transition at moderate energies. In fact, they increase
from $\sim$3~km~s$^{-1}$ in the low energy transitions to $\sim$7~km~s$^{-1}$ 
in transitions with E$_u$$>$50~K. But this trend is not valid for higher energy lines which 
seem to have a constant linewidth of $\sim$4~km~s$^{-1}$. 
A similar behavior is found when one compares the
velocity of the line with the upper state energy. The line velocity changes from
v$\sim$$-$10~km~s$^{-1}$ to $\sim$$-$7.5~km~s$^{-1}$ when the energy increases
from $\sim$10~K to 50~K. However, for higher energies, the line velocities seem to
make the reverse way and change from $\sim$$-$8~km~s$^{-1}$ to $-$10~km~s$^{-1}$.
This suggests that the emission of  the low and moderate energy transitions ( E$_u$$>$50~K)  arise 
in the molecular outflow. The correlation found between the linewidth and the energy of the
transition suggests that the high velocity gas is associated with higher excitation temperatures.
This result is consistent with the density estimated from the SiO lines. 
For E$_u$$>$ 50 K, there is a jump in the line velocity which
go back to $\sim$ -10 km s$^{-1}$ . We propose that this could be due to
the existence of a hot component 
in the CH$_3$OH emission in addition to that related to the bipolar outflow. 

In Fig.~11 we show the CH$_3$OH rotational
diagram for FIRS~2. All the observed CH$_3$OH transitions cannot
be fitted with a single straight line. We need to assume at least two rotation
temperatures to fit the observational data.
The low energy transitions (E$_u$$<$50 K) are well fit with a T$_{rot}$$\sim$17 K,
while the high energy transitions require a higher rotation temperature, T$_{rot}$$>$80 K.
We propose the existence of a $``$hot core" component which dominates the
emission in the high energy transitions.

Several CH$_3$OH lines have also been observed in LkH$\alpha$ 234. In this case all the lines
are centered at the ambient velocity. However, there are important variations in the linewidths of the
observed lines. Like in the case of FIRS~2, the linewidths seem to increase with the
energy of the upper level for E$_u$$<$50 K (see Fig. 12).  We have only detected the methanol lines towards
the (0,0) position and consequently, we have no information about the size of the
emitting region. Thus, we have considered the
two limiting cases of a point source and a beam filling factor of 1 to make the
rotational diagram. In the first case, we need two
gas components to fit all the observed transitions. The cold one would have T$_{rot}$$\sim$24 K,
and the hot one, T$_{rot}$$>$250 K. But if we assume a beam filling factor of 1, all the oberved
transitions are well fit with a T$_{rot}$$\sim$60 K. With our data, we cannot discern
between these two cases.

\setlength\unitlength{1cm}
\begin{figure}
\vspace{13cm}
\includegraphics{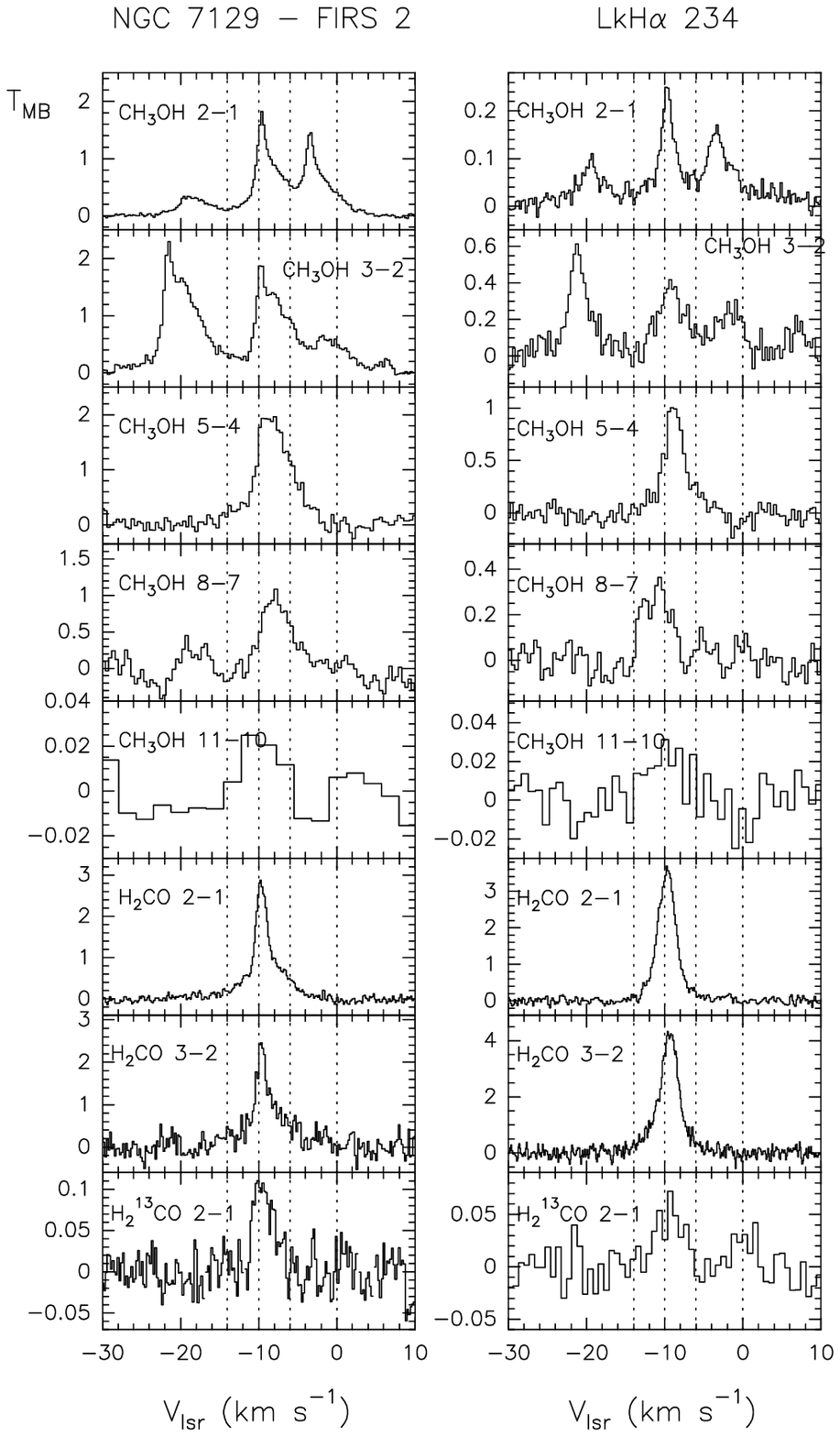}
%\end{picture}
\caption{Observed spectra of CH$_3$OH and H$_2$CO lines towards the
(0,0) position in NGC~7129~--~FIRS~2 and LkH$\alpha$~234..}
\end{figure}

\setlength\unitlength{1cm}
\begin{figure}
\vspace{8cm}
\includegraphics{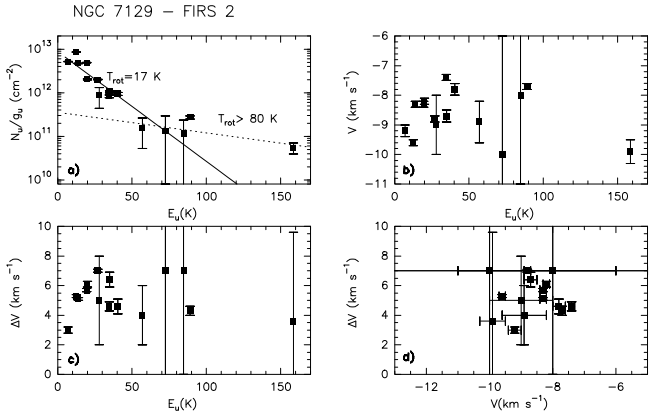}
%\end{picture}
\caption{The (a) panel shows the rotational diagram for CH$_3$OH in 
the (0,0) position of NGC~7129~--~FIRS~2. We have degraded the angular
resolution of the 1.3mm maps (the 5$\rightarrow$4 and 8$\rightarrow$7 methanol lines) in order to 
have the angular resolution of the 3mm maps (lines 2$\rightarrow$1 and 11$\rightarrow$10 lines).
Two rotation temperatures (T$_{rot}$=17~K and T$_{rot}$$>$80~K are required
to fit the data).
In the (b) and (c) panels we plot the line velocites and linewidths, 
V and $\Delta$V, of the 
CH$_3$OH lines as given by the gaussian fits shown in Table 3 versus the
upper state energy of the observed transition (E$_u$). Finally, the (d) panel
shows V versus $\Delta$v for the CH$_3$OH lines. This panel clearly shows the
existence of two veloctiy components, one at $\sim$$-$10~kms$^{-1}$ and the other 
at $\sim$$-$6~kms$^{-1}$ in this position.}
\end{figure}

\setlength\unitlength{1cm}
\begin{figure}
\vspace{8cm}
\includegraphics{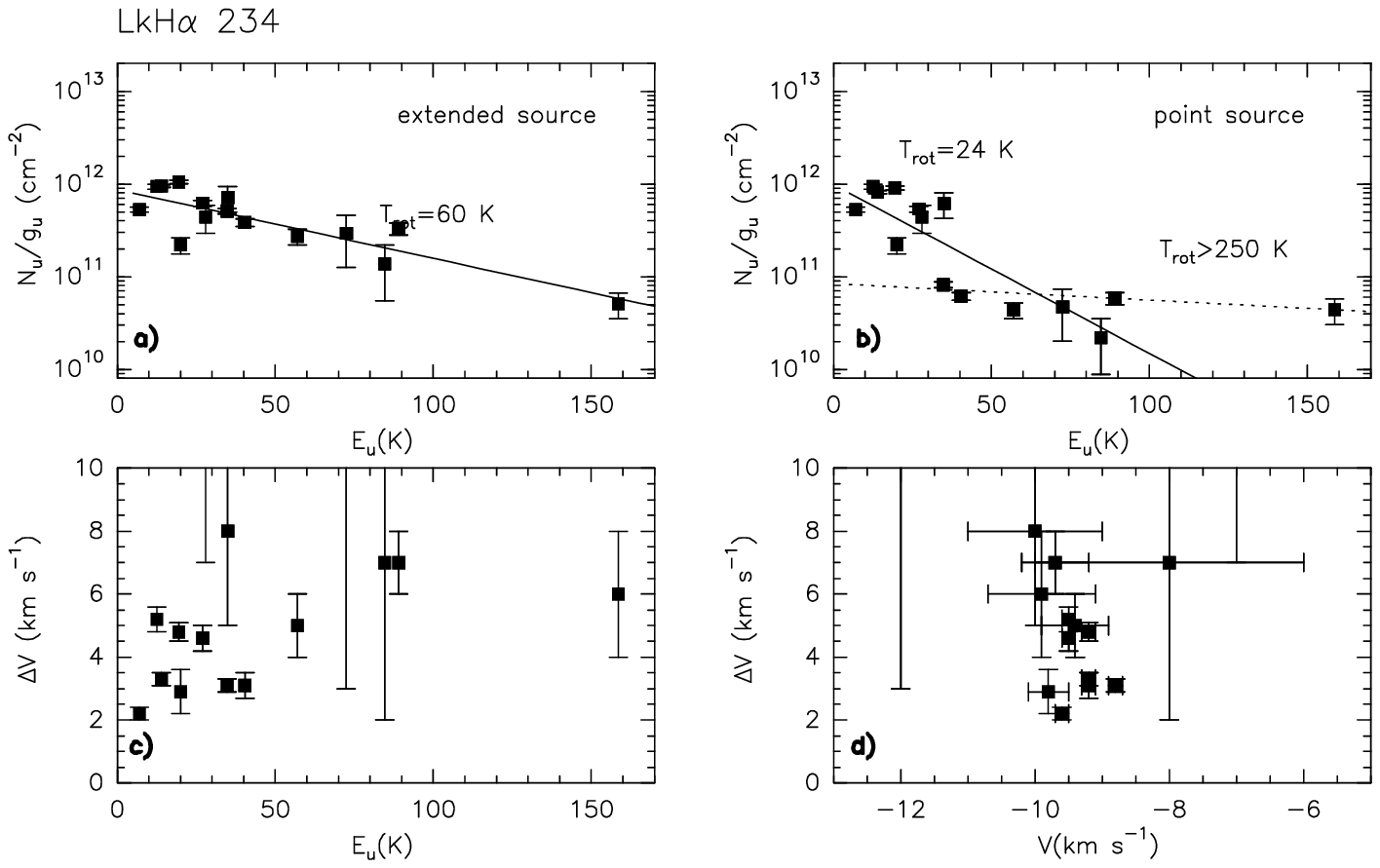}
%\end{picture}
\caption{The top panels show the rotational diagram for CH$_3$OH in LkH$\alpha$~234
in the two limit cases assumed in the paper, panel (a) corresponds to
a beam filling factor $\sim$ 1and panel (b) corresponds to a point source. In
the bottom panels we plot the linewidths, $\Delta v$,  derived from the Gaussian fits versus the upper
energy state of the transition, E$_u$, and the central velocity, V, for all the lines. In the case
of LkH$\alpha$~234 we have only one velocity component.}
\end{figure}
		
\subsection{H$_2$CO}

We have observed two H$_2$CO rotational transitions and one of the rarer isotope H$_2$$^{13}$CO$^+$
toward FIRS~2 and LkH$\alpha$~234. The obtained spectra are shown in Fig.~10.
Similarly to other molecules, the H$_2$CO spectra towards FIRS~2 present two well
differentiated components, a narrow one centered at $\sim$$-$9.6 km s$^{-1}$ with a linewidth of
$\sim$1.3~kms~s$^{-1}$ and a much wider one centered at $\sim$$-$9.0~km~s$^{-1}$. However, the wide
component has not the typical R1 profile observed in the SiO, CS and methanol lines at the star
position. While the profiles of the SiO,CS and methanol lines present only red wings, the H$_2$CO lines
present a quite symmetric profile with blue and red wings. Consequently,
the central velocity of the wide H$_2$CO component is similar to that of the ambient gas
and the linewidth is as large as $\Delta$v=7 km s$^{-1}$. 
To further investigate the nature of these components we have studied the integrated intensity maps of 
the H$_2$CO  2$_{12}$$\rightarrow$1$_{11}$ line for the different velocity intervals (see Fig. 13). 
The most striking feature could be the jet-like morphology observed in the H$_2$CO emission at
blue velocities (from -15 to -10 km s$^{-1}$). At red velocities, the emission is maximum at the 
offset (-7$''$,-7$''$) which is located close to the bullet R1.
Since the wide component have a very well differentiated profile, we have
been able to subtract the wide component to the observed spectra and mapped both
components separately. Our results are quite suggestive. The wide component presents a jet-like
morphology with the maximum towards the position R1. The morphology of the narrow component is an
intense ridge which surrounds the jet. This strongly suggests that the narrow component is tracing
the shocked gas of the molecular cloud which is interacting with the jet. 
The maximum of this narrow component coincides with the position where the bullet R1
impinges in the cloud.
We have derived rotation temperatures and column densities in the narrow and wide components
separately, and obtained similar excitation conditions in both components. 
Thus, although the kinematics is clearly different, the physical
condition of both components are quite similar. 

In Fig.~10 we show the H$_2$CO spectra towards LkH$\alpha$ 234. The profiles of the
H$_2$CO lines in this source, also suggest the existence of a narrow and wide
components. However, these two component cannot be easily separated.
For this reason, we have derived rotation temperatures 
by considering the sum of the two components. We obtain a rotation temperature and H$_2$CO column
density similar to those obtained in FIRS~2. In Fig. 5 we show the integrated
line intensity maps in this source. Similarly to the case of FIRS~2 we find emission
along the outflow and in a direction perpendicular to it. Thus far, no bipolar outflow has been 
detected in this direction. Thus, this H$_2$CO emission is associated with the flattened clump in
which the Herbig Be star LkH$\alpha$ 234 is embebded that is being heated by the recently born
star.  		

\begin{figure*}
\vspace{8cm}
\includegraphics{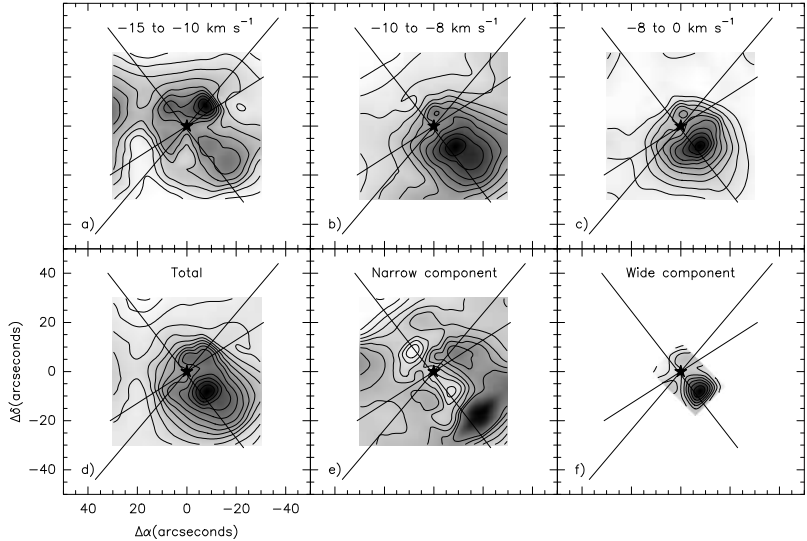}
%\end{picture}
\caption{ Panels a-d are the integrated intensity maps of the H$_2$CO 2$_{12}$$\rightarrow$1$_{11}$ line towards 
NGC~7129~--~FIRS~2 for different velocity intervals. 
In panel e) we show the integrated intensity map of the H$_2$CO lines after subtracting the
Gaussian fit to the wide component from all the spectra. In panel f) we show the area of this
wide component as derived from the Gaussian fit. The axis of the outflow FIRS~2-out~1 and FIRS~2-out~2
are indicated.
Contour levels are: a) 0.5 to 5 by 0.5 K kms$^{-1}$;
b)1.0 to 6.5 by 0.5 K kms$^{-1}$; c) 0.5 to 5 by 0.5 K kms$^{-1}$; d)1 to 14 by 1 K kms$^{-1}$;
e) 0.5 to 5 by 0.5 K kms$^{-1}$; f) 1 to 11.5 by 1 K km s$^{-1}$.}
\end{figure*}

\subsection{CH$_3$CN}

We have observed the CH$_3$CN 5$\rightarrow$4 and 13$\rightarrow$12 lines towards
FIRS~2 and LkH$\alpha$~234. Because
of the rotational structure of CH$_3$CN, one can observe several 
lines at different energies very close 
in frequency. This allows us to estimate the rotation temperature avoiding observational errors 
and the uncertainty due to the unknown source size. 
We have carried out these calculations towards our two sources. Unlike the other molecules
observed, we do not detect a cold component in the CH$_3$CN lines, but only the
warm one. The detection of
a hot CH$_3$CN component with T$_k$ $>$ 63 K in FIRS 2 shows 
the existence of a hot core in this object. 

CH$_3$CN seems to be the best tracer of hot cores in these intermediate-mass stars.
Contrary to CH·$_3$OH and H$_2$CO whose low energy lines arise mainly in the bipolar
outflow, the rotational lines of CH$_3$CN seems to arise in the hot core and 
provide a good measure of the kinetic temperature
of this hot component.
		
\setlength\unitlength{1cm}
\begin{figure}
\vspace{5cm}
\includegraphics{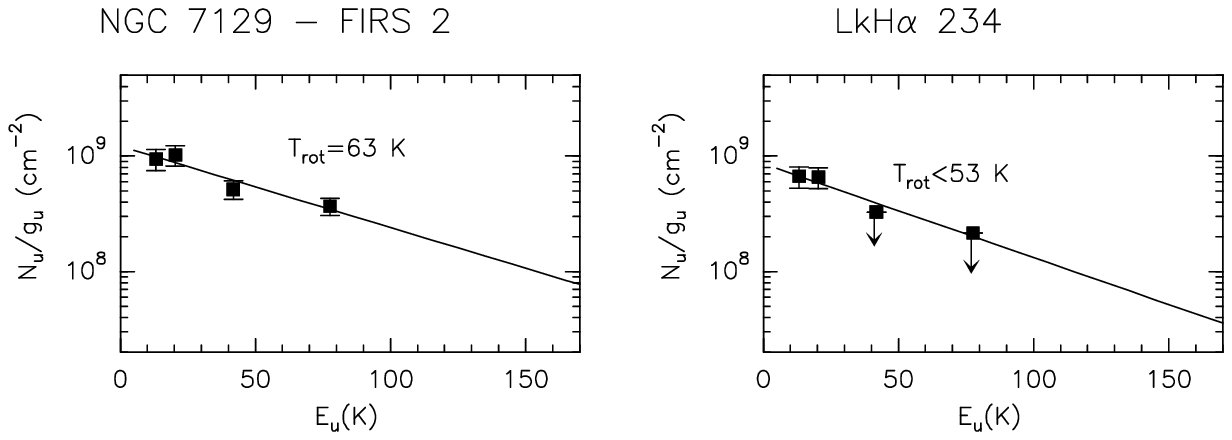}
%\end{picture}
\caption{Rotational diagram of CH$_3$CN in the (0,0) position of FIRS~2
and LkH$\alpha$~234.  }
\end{figure}

\subsection{CN,HCN}
We have mapped a small region around FIRS~2 and LkH$\alpha$ 234 in the HCN~1$\rightarrow$0 and 
CN~1$\rightarrow$0 and 2$\rightarrow$1 lines. The integrated intensity maps around 
FIRS~2 are shown in Fig.~4 and 5.  Thus far, we have two different kind of molecules
depending on the morphology of their emission in FIRS~2. The first group is formed by the molecular ions N$_2$H$^+$ 
and H$^{13}$CO$^+$ which peak at the YSO position and trace the bulk of the cold envelope. The second group is formed by CS, SiO,
CH$_3$OH and H$_2$CO peaks at the position R1 and its emission has an important contribution from the molecular outflow
FIRS~2-out~1. The morphology of the CN~1$\rightarrow$0 emission is different from those of these two groups. The CN emission
does not peak at the star position nor at the position of the bullet R1, but to the north, suggesting a new chemical differentiation 
in the protostellar envelope.  The morphology of the HCN emission is also different from those of the other molecular lines.
In fact, the HCN emission peaks to the NW from the star position, in a position intermediate between the bullet R1 and the peak
of the CN emission. The large linewidths of the HCN~1$\rightarrow$0 line (see Fig.~15) show that, 
in contrast to the CN~1$\rightarrow$0 line, the
emission of the HCN~1$\rightarrow$0 line has a significant contribution from the molecular outflow giving rise to this 
peculiar morphology. 

The radicals CN and HCN are known to be specially abundant in PDRs. In particular, the CN/HCN ratio has been successfully
used as a PDR tracer in different kind of objects. In Fig.~16,we show the maps 
of the (CN 1$\rightarrow$0)/(HCN 1$\rightarrow$0) intensity ratio in
FIRS~2 and LkH$\alpha$~234. The (CN 1$\rightarrow$0)/(HCN 1$\rightarrow$0) line intensity ratio is maximum at the star 
position and to the north, forming a conical feature with the star at its apex. 
We have been estimated the CN rotation temperature 
from the (CN 2$\rightarrow$1)/(CN 1$\rightarrow$0) line intensity ratio (see Table~4).
Assuming the LTE approximation and the same rotation temperature for CN and HCN, 
we obtain a CN/HCN abundance ratio of $\sim$3 at the star position. 
This value is similar to those found in PDRs and suggests that the gas chemistry in this conical feature 
is being affected by the UV radiation from
the protostar. However, the axis of this conical feature seems to be more similar to that of the
outflow FIRS~2-out~2 than to that of the outflow FIRS~2-out~1. This suggests that the PDR traced by the
high CN/HCN ratio could be related to the star driving the outflow FIRS~2-out~2 instead of to the
Class 0 object. In fact, the PDR could be formed in the walls of the cavity excavated by the 
outflow FIRS~2-out~2 when they are illuminated by the exciting star. But observations with
higher spatial resolution are required to conclude about this point.

We have also observed the CN and HCN lines towards LkH$\alpha$~234. 
In this case, the linewidths of the CN and HCN lines are similar, 
and in agreement with those found in the warm component. 
But, contrary to most of the observed molecular species, the 
CN and HCN emission do not peak at the star position but to the north forming
a conical feature. We have calculated the CN/HCN integrated intensity 
ratio in the region. Surprisingly, 
the CN/HCN ratio is minimum at the star position and maximum at the border 
of the clump as traced by the 1.3mm observations suggesting that the clump is illuminated 
from outside. Making column densities estimates, 
we derive N(CN)/N(HCN)$\sim$3 at the star position.

Thus, the CN/HCN fractional abundance ratio at the star position is equal (within the uncertainties) 
in FIRS~2 and LkH$\alpha$~234 and consistent with the
expected value in a PDR. However the behavior of the CN/HCN ratio is very different in the rest of the envelope. 
In the case of FIRS~2, the
CN/HCN ratio decreases outwards from the star as expected from a PDR illuminated
from the interior and with an optically thick envelope.
In the case of LkH$\alpha$~234, the CN/HCN ratio is quite constant inside the clump and increases at the edges. 
This suggests that the clumps is also illuminated from outside (the clump is located at the border of an HII region). 
Since the envelope is less massive than that associated with FIRS~2, 
the whole envelope can be considered as a PDR.
		
\setlength\unitlength{1cm}
\begin{figure}
\vspace{14cm}
\includegraphics{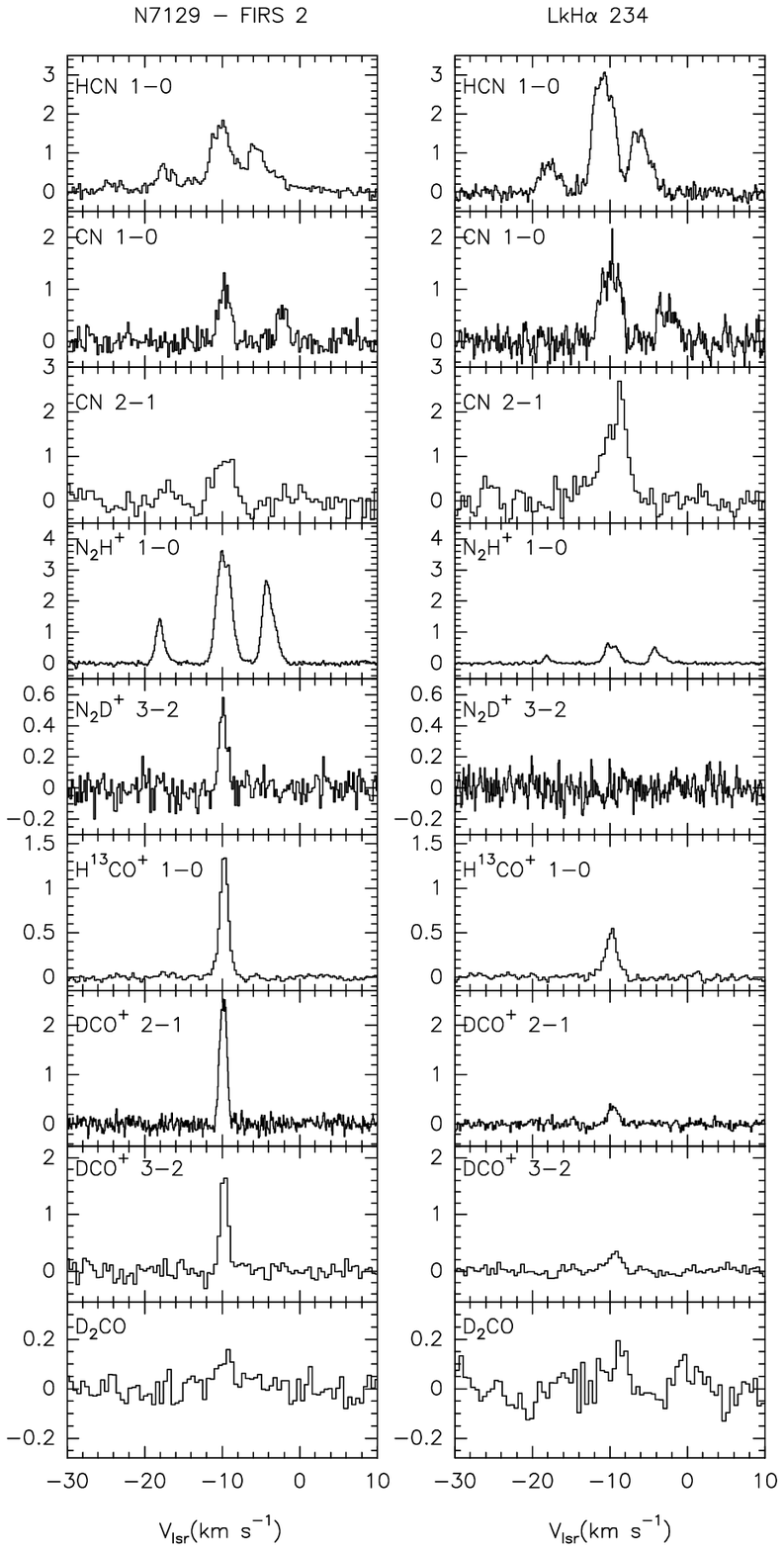}
%\end{picture}
\caption{Observed spectra towards the (0,0) position in LkH$\alpha$~234 and NGC~7129~--~FIRS~2.
The warming of the gas
during the protostellar evolution, produces significant chemical changes specially in the
cold component.}
\end{figure}
		
\setlength\unitlength{1cm}
\begin{figure}
\vspace{6cm}
\includegraphics{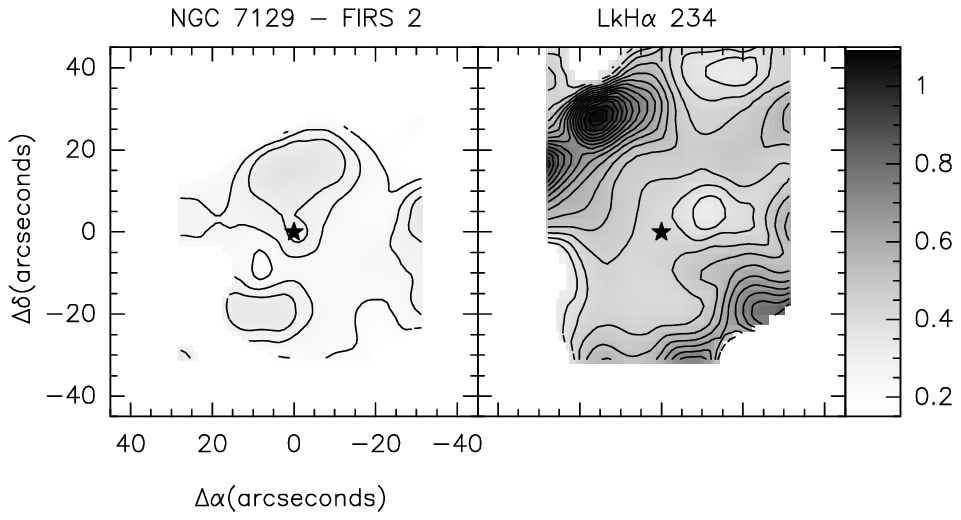}
%\end{picture}
\caption{CN 1$\rightarrow$0 /HCN 1$\rightarrow$0 integrated intensity ratio in NGC~7129~--~FIRS~2 (left)
and LkH$\alpha$~234 (right). Contour levels are 0.10 to 0.35 by 0.05 in NGC~7129~--~FIRS~2 and 
0.35 to 1.1 by 0.05 in LkH$\alpha$~234.}
\end{figure}

\section{Deuterated compounds}

In order to derive the deuterium fractionation we have observed the DCO$^+$ 2$\rightarrow$1 
and 3$\rightarrow$2,  N$_2$D$^+$ 3$\rightarrow$2 , and D$_2$CO 4$_{04}$$\rightarrow$3$_{03}$ lines
toward the studied regions. 
In Fig.~15 we show the spectra of these lines toward FIRS~2 and LkH$\alpha$~234.
The integrated line intensity maps of the DCO$^+$ 2$\rightarrow$1 
line toward FIRS~2 are shown in Fig.~4.  

The linewidths of the DCO$^+$ and N$_2$D$^+$ lines are 
$\sim$1.0 km s$^{-1}$ suggesting that they arise in the cold component of the envelope like
the non-deuterared compounds HCO$^+$ and N$_2$H$^+$. 
The linewidth of the D$_2$CO line is $\sim$4 km s$^{-1}$ like those
of the lines arising in the warm component, and in particular, the lines of the
chemically related species 
H$_2$CO and H$_2$$^{13}$CO. In Table 9 we show the DCO$^+$/H$^{13}$CO$^+$,
N$_2$D$^+$/NH$_2$$^+$ and D$_2$CO/H$_2$CO abundance
ratios in both sources. 
The DCO$^+$/H$^{13}$CO$^+$ abundance ratio is a factor of 20 lower in
LkH$\alpha$ 234 than in FIRS~2.  This factor is so large that cannot
be due to the H$^{13}$CO$^+$ depletion but to a different value of the deuterium
fractionation in these cold envelopes. Thus, we propose that the deuterium
fractionation in the cold envelope decrease during the protostellar evolution.
As we will comment in detail in next section, this increase in the deuterium 
fractionation can be understood as the consequence of the envelope warming during the
protostellar evolution.  

A very different case is the D$_2$CO/H$_2$CO abundance ratio which increases by
a factor $\sim$1.5 from FIRS~2 to LkH$\alpha$~234.  Since a factor 
of 1.5 is within the uncertainities of our column density estimates, we conclude
that the deuterium fractionation, as measured by the D$_2$CO/H$_2$CO abundance
ration, seems to be  constant (or slightly increase) in the warm component
during the protostellar evolution. Thus, the evolution of the deuterium fractionation in
the warm envelope seems to follow a different trend than in the cold envelope. The
evaporation of the icy grain mantles is very likely the main responsible of this behavior.
		
\setlength\unitlength{1cm}
\begin{figure*}
\vspace{13cm}
\includegraphics{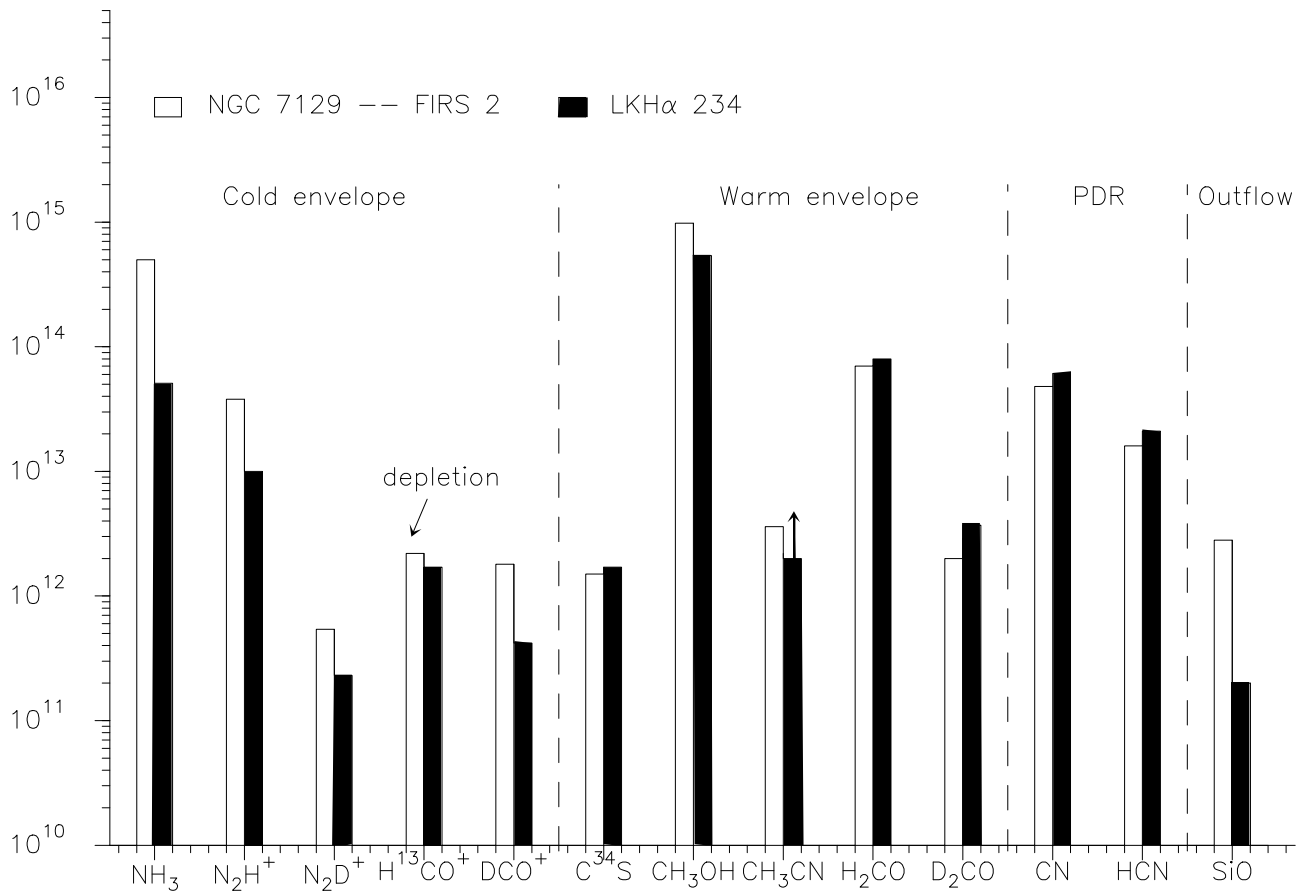}
%\end{picture}
\caption{Histogram with the total beam-averaged molecular column densities estimated in
the IM Class 0 YSO NGC 7129 -- FIRS 2 and the HBe star LkH$\alpha$ 234.
The column densities
of the molecules tracing the cold envelope decrease at least by a factor of 5 between NGC~7129--FIRS~2
and LkH$\alpha$234. An exception is H$^{13}$CO$^+$ that is depleted in the Class 0 protostar (see text). The
column densities of the molecules tracing the warm envelope and the PDR remain quite constant, in some cases
increase, from the Class 0 protostar to the HBe star. The SiO column density is lower by almost 2 
orders of magnitude in LkH$\alpha$ 234 than in NGC~7129--FIRS~2. This is very likely related to the absence
of an energetic bipolar outflow in the HBe star.}
\end{figure*}

\section{Discussion}

\subsection{Physical structure of the YSOs envelopes and its evolution}
Our data show the existence of at least two well differentiated components in the envelope
of FIRS~2 and LkH$\alpha$~234, the cold envelope traced by the 
low energy lines of NH$_3$, N$_2$H$^+$ and H$^{13}$CO$^+$, and the warm 
envelope traced by the CS, CH$_3$OH and H$_2$CO lines. 
These two components can be differentiated 
observationally by their kinematics, the morphology of their emission and by their physical conditions.
Thus the lines arising in the cold component are narrow ($\Delta$v$\sim$1 km s$^{-1}$)
and the emission peak is located at the star position in FIRS~2. Besides,
the kinetic temperature of this gas estimated from the NH$_3$ lines is $\sim$13 K
in FIRS~2 and $\sim$28 K in LkH$\alpha$ 234. 

A warm envelope component is detected towards these sources traced by the
emission of species like CS, CH$_3$OH and H$_2$CO. These species present 
enhanced abundances in regions where the icy grain mantles are evaporated
\citep{van00}.
They are also abundant in molecular ouflows where they can be released to the
gas phase by shock fronts. In FIRS~2, the emission of the low energy
transitions of these species arise mainly in the bipolar outflow.
However, this association is not clear in LkH$\alpha$~234 where their
emisison could arise in the inner and warmer part of the envelope.

Finally, we have strong evidences for the existence of a hot core in
the Class 0 protostar FIRS~2. The high density measured at ambient
velocities from the C$^{34}$S lines, the high temperature component of the 
CH$_3$OH lines and, above all,  the detection of the CH$_3$CN lines with 
rotation temperature of $\sim$63~K show the existence of a hot core in
this target.
  
\subsection{Physical and Chemical evolution of the YSOs envelopes}
\subsubsection {Cold envelope}
 
In Table~9, we show the physical paramenters and molecular column densities in
the Class 0 protostar FIRS~2 and LkH$\alpha$~234.
The molecules NH$_3$, N$_2$H$^+$, H$^{13}$CO$^+$ and their deuterated compounds
DCO$^+$ and N$_2$D$^+$ trace the cold envelope component. The column
densities of these species decrease by a factor of 5--10 from the Class 0 protostar to
the Type I Herbig Be star showing that the mass of the cold envelope decrease
by at least a factor of 5 during the protostellar phase.
Based on the NH$_3$ and N$_2$H$^+$ data we have also derived
the kinetic temperature and size of this cold component. The kinetic temperature increases 
from $\sim$13 to $\sim$28 K and the size of the emitting region decreases from 21$''$ to 6$''$ from 
FIRS~2 and LkH$\alpha$~234.  
This is consistent with previous results by \citet{fue02} which shows the protostellar
envelope is dispersed and becomes warmer during its evolution to become a visible star. 

The warming of the cold envelope produces changes in its chemical composition.
In Table 5 we show the N$_2$H$^+$/H$^{13}$CO$^+$ and NH$_3$/N$_2$H$^+$ abundance
ratios in several positions. Note that the N$_2$H$^+$/H$^{13}$CO$^+$ ratio 
is different in the studied objects. The N$_2$H$^+$/H$^{13}$CO$^+$ ratio is $\sim$17 in the cold 
young object FIRS~2 and $\sim$6 in the more evolved and warmer Herbig Be star 
LkH$\alpha$~234. As discussed in Section 3.2, this gradient in the N$_2$H$^+$/H$^{13}$CO$^+$
ratio is very likely due to the H$^{13}$CO$^+$ depletion in the cold envelope of the protostar.
Molecular  depletion is expected to be significant only for  kinetic temperatures T$_k$$<$20 K. 
Thus, molecular depletion is negligible in the envelope of LkH$\alpha$ 234
where the gas kinetic temperature is $\sim$28 K.

Within the nitrogen chemistry, we have also studied the NH$_3$/N$_2$H$^+$ ratio. Since the
beam is very different for the NH$_3$ and N$_2$H$^+$ observations, this abundance ratio is very dependent on
the assumed source size. For LkH$\alpha$ 234 we have been able to calculate both,
the NH$_3$ and N$_2$H$^+$ emitting regions sizes. We have derived the same size, $\sim$ 6$''$--8$''$, 
for both molecules
and the NH$_3$/N$_2$H$^+$ ratio is $\sim$10. In the case of FIRS~2  we have not been able to derive
the source size from the NH$_3$ emission but we have estimated a size of $\sim$21$''$ from the
N$_2$H$^+$ observations. We have assumed two limiting cases for the calculations of the NH$_3$/N$_2$H$^+$ ratio
in this source.  Assuming
a beam filling factor of $\sim$1 for both molecules we obtain   NH$_3$/N$_2$H$^+$$\sim$13 in this object.
However if we assume that the size of the NH$_3$ emission is $\sim$ 21$''$, like in the case of N$_2$H$^+$,
we obtain NH$_3$/N$_2$H$^+$$\sim$33. In this case, we would have an NH$_3$ abundance enhancement
in the colder envelope of the Class 0 protostar. Recent results in pre-stellar core show that the NH$_3$ abundance
could be enhanced in dense regions of these cores where the CO is expected to be depleted \citep{taf04}.

\subsubsection{Warm envelope} 
 
When the star heats the envelope a sublimation front proceeds outwards the star removing molecules from grain mantles. 
The region of the envelope in which the gas kinetic temperature is high enough to evaporate the grain mantles is
what we have called $``$warm envelope''. The species 
released to gas phase are called $``$parent molecules" and increase significantly their abundances. The molecules CH$_3$OH,
NH$_3$, and H$_2$CO are within this group. These molecules drive a high temperature chemistry giving rise to $``$daughter" molecules
like CH$_3$CN. Within this scheme, CH$_3$OH, H$_2$CO and CH$_3$CN are tracers of the warm part of the envelope where the
ices have been evaporated \citep{rod03}. 
Some of these species are also abundant in the molecular outflow where shock fronts remove
them from the grains mantles, and to a lesser extent in the cold envelope. This is the case for NH$_3$ with the emission of the low-lying
transitions dominated by the cold envelope in both,  FIRS~2 and LkH$\alpha$ 234, while the (3,3) line arise in the 
warmer component. 

In Table 9 we show the physical conditions and the molecular column densities in the warm envelope
of FIRS~2 and LkH$\alpha$~234. Contrary to the species tracing the cold envelope, the species
tracing the warm envelope present similar column densities in both targets (see Fig.~17). In fact, some species like
H$_2$CO and C$^{34}$S seem to have larger column densities in LkH$\alpha$~234.
On the other hand, the size
of the warm envelope is larger in the case of the Herbig Be star than in 
the case of the Class 0 YSO. Thus, the mass and size of the
warm envelope remains quite constant, or even increase, during the protostellar evolution.

Although the column densities of the $``$warm envelope'' species are not very different
in the two YSOs, the origin of their emission could be different.
The CH$_3$OH and H$_2$CO emission seem to be dominated by the molecular  in FIRS~2. 
In the case of CH$_3$OH we have detected two components in FIRS~2. The one centered at $\sim$7 km s$^{-1}$
is clearly associated with the molecular outflow FIRS~2-out~1 and dominates the CH$_3$OH emission in all the
transitions with  E$_u$$<$100~K. We have observed only two low-energy H$_2$CO lines in our targets. The emission
of these lines in FIRS~2 is clearly associated with the outflow FIRS 2-out 1 (see Fig.~13). 
   
Intense CH$_3$OH and H$_2$CO lines have also been detected in LkH$\alpha$ 234. 
In this case the link between these species and the
molecular outflow is not clear. In fact, these lines could arise from a warm inner envelope 
where the icy mantles are being evaporated releasing
these species to the gas phase. Thus, although the CH$_3$OH and H$_2$CO column 
densities are very similar in both objects,
the mechanism which removes these species from the icy mantles can be different in the Class 0 
protostar FIRS 2 and the Herbig Be star LkH$\alpha$ 234. While shock fronts are very likely
the main mechanism for the erosion of grain mantles in FIRS~2, 
ice evaporation could be a dominant mechanism in LkH$\alpha$~234.

\subsubsection{CN,HCN}

The CN and HCN column densities  increases from FIRS~2 to LkH$\alpha$~234
by a factor of $\sim$ 6. This is easily understood within an evolutionary trend.
Since the cold envelope has already been dispersed in LkH$\alpha$~234, the UV radiation can 
penetrate deeper into the cloud affecting the chemistry all across the envelope 
in LkH$\alpha$~234. The enhancement of ionization fraction (in particular, the enhancement in
the C$^+$ abundance) in the LkH$\alpha$ 234 envelope
could produce an enhancement in the fractional abundances of the nitrogenated chains
CN and HCN relative to N$_2$H$^+$ and NH$_3$.

The CN/HCN ratio has been widely used as a PDR tracer thus far. Values of the CN/HCN ratio $>$1
has been considered as a prove of the existence of a photon-dominated chemistry region
\citep[see e.g.][]{fue93,fue95,fue03,bac97,rod98}.  
We have detected a small region in the vicinity of the Class 0 source in FIRS~2 in
which the CN/HCN ratio is $>$1. This strongly suggests the existence of an incipient PDR in the
inner part of this protostellar envelope. However, our limited angular resolution prevent us to
discern if this PDR is associated with the Class 0 source or with the IR companion star which
is driving the outflow FIRS~2-out~2. 

\subsubsection{Deuterated compounds}

We have observed DCO$^+$ and N$_2$D$^+$ in order to be able to derive the D/H ratio in the cold envelope of
these stars and its possible changes during the central object evolution. 
Enhancements of the D/H ratio over the 10$^{-5}$
ratio of [HD]/[H$_2$] have been found in dark clouds and young protostars \citep[e.g.][]{but95,
wil98}. There are two main ways of producing these 
enhancements. Firstly, grain surface chemistry may enhance molecular D/H ratios \citep{tie83,bro89a,bro89b,cha97}. 
Secondly, some key gas phase reactions involving destruction of deuterated species run slower at low temperatures
than the equivalent reactions with hydrogen, and this leads to molecular D/H enhancements where a cold gas phase chemistry
has been active \citep{rob00a,rob00b}. 
Furthermore, in colder gas, depletion of heavy molecules such as CO results in an increase of 
[H$_2$D$^+$]/[H$_3^+$] and molecular D/H ratios \citep{dal84,bro89a,rob00a,rob00b}.
We have found DCO$^+$/H$^{13}$CO$^+$$\sim$0.7 in FIRS~2 and $\sim$0.25 in LkH$\alpha$ 234. Assuming a
$^{12}$C/$^{13}$C isotopic ratio of 89, this would imply DCO$^+$/HCO$^+$$\sim$0.008 in FIRS~2 and 
$\sim$0.003 in LkH$\alpha$ 234. These ratios are similar to that found in low mass Class 0 protostar 
IRAS+16293--2422 by \citet{van95} but are lower that the values found in dark clouds 
\citep[see e.g.][]{tin00}. A cold gas chemistry as well as depletion of heavy
molecules can explain the enhanced values DCO$^+$ /HCO$^+$ ratio found in the young stellar object 
FIRS~2.

We have also observed the doubly deuterated formaldehyde D$_2$CO. The linewidths of the observed
D$_2$CO lines show that this species, like the non-deuterated compound H$_2$CO, arise in the warm envelope. 
The D$_2$CO column densities as well as the D$_2$CO/H$_2$CO ratio, are similar in LkH$\alpha$ 234
than in FIRS 2. This suggests that the deuterium fractionation remains constant (or even
increase, see Table 9) in the warm envelope during the protostellar evolution. Thus, the
evolution of the deuterium fractionation is different in the cold and warm part of the envelope.
Grain-surface chemistry may enhanced the deuterium fractionation in the warm envelope 
where molecular evaporation is very likely the main chemical phenomenum, while the
cold gas chemistry and depletion could determine the evolution of the deuterium fractionation in
the cold envelope. 

The deuterium fractionation has been proposed as a chemical clock
in YSOs. Our data confirms that the deuterium fractionation changes significantly during the
protostellar evolution and consequently, can be used as a chemical clock. However, these changes 
are different in the cold and in the warm part of the envelope, and  the measured D/H ratio
is very dependent on the molecular compounds used to determine it.

\begin{table}[]
\caption{Beam averaged column densities}
\begin{tabular}{lcc}
\\ \hline
\multicolumn{1}{c}{} & 
\multicolumn{1}{c}{NGC 7129 -- FIRS 2} &       
\multicolumn{1}{c}{LkH$\alpha$ 234} \\
Age           &  $\gsim$ 3000 yr              & $\sim$ 10$^5$ yr \\ 
n(cm$^{-3}$)   & 5 10$^5$ -- 3 10$^6$   &  2 10$^5$ \\  
\hline
\multicolumn{3}{c}{Cold envelope} \\
T$_k$                 &     13 K               &  28 K \\
$\Omega_s$        &    $\sim$ 21 $''$   &   $\sim$ 8$''$   \\
NH$_3$              &    4.9 10$^{14}$    &   4.0 10$^{13}$ \\
N$_2$H$^+$       &    3.8 10$^{13}$     &  1.0 10$^{13}$  \\
H$^{13}$CO$^+$  &   2.2 10$^{12}$     &  1.7 10$^{12}$  \\
DCO$^+$             &  1.8 10$^{12}$     & 4.2 10$^{11}$   \\
N$_2$D$^+$        &   5.4 10$^{11}$     &   $<$ 2.3 10$^{11}$ \\
N$_2$H$^+$/H$^{13}$CO$^+$    &   17   &    6    \\
NH$_3$/N$_2$H$^+$                 &  13 -33$^b$  &  10  \\
DCO$^+$/H$^{13}$CO$^+$         &   0.7      & 0.2   \\
N$_2$D$^+$/N$_2$H$^+$          &    0.014  & $<$ 0.02 \\  \hline
\multicolumn{3}{c}{Warm envelope} \\
T$_k$                &  $>$ 50 K      &  $>$ 100 K \\
$\Omega_s$       &  $\sim$ 9$''$  & $\sim$ 14$''$ \\
C$^{34}$S          &  1.5 10$^{12}$   &  1.7 10$^{12}$  \\
CH$_3$OH         &  9.8 10$^{14}$   &  5.4 10$^{14}$  \\
CH$_3$CN          &  3.6 10$^{12}$   &  $>$ 2.0 10$^{12}$ \\
H$_2$CO           &  7.0 10$^{13}$    &  8.0 10$^{13}$  \\
H$_2$$^{13}$CO  & 2.3 10$^{12}$    & 2.4 10$^{12}$  \\
D$_2$CO             & 2.0 10$^{12}$    &  3.7 10$^{12}$  \\
D$_2$CO/H$_2$CO  & 0.03              &  0.05    \\  \hline
\multicolumn{3}{c}{PDR} \\
CN                   &   4.8 10$^{13}$       & 6.1 10$^{13}$   \\
HCN                 &   1.6 10$^{13}$       & 2.1 10$^{13}$   \\ \hline
\multicolumn{3}{c}{Outflow} \\
SiO                  &   2.8 10$^{12}$        & 2.0 10$^{11}$   \\ \hline
\end{tabular}

\noindent
$^b$ Assuming beam filling factor $\sim$ 1 and a source size of 21$''$ for  the NH$_3$ emitting region.
\end{table}

\begin{table}[]
\caption{Chemical diagnostics for YSO evolution}
\begin{tabular}{lcc}
\\ \hline
\multicolumn{1}{c}{} & 
\multicolumn{1}{c}{NGC 7129 -- FIRS 2} &       
\multicolumn{1}{c}{LkH$\alpha$ 234} \\
\hline
%N$_2$H$^+$       &    1     &  1  \\
%NH$_3$              &    $\sim$ 1 10$^{-9}$      &  $\sim$ 1 10$^{-9}$ \\
%H$^{13}$CO$^+$  &   $\sim$ 6 10$^{-12}$     &  $\sim$ 2 10$^{-11}$  \\
%C$^{34}$S           &    $\sim$ 4 10$^{-12}$     &  $\sim$ 3 10$^{-11}$  \\
%CH$_3$OH         &    $\sim$  3 10$^{-9}$       &  $\sim$ 5 10$^{-9}$  \\
%CH$_3$CN          &   $\sim$  1 10$^{-11}$     &  $>$ 10$^{-11}$ \\
%H$_2$CO           &    $\sim$  2 10$^{-10}$     &  $\sim$ 8 10$^{-10}$  \\
%CN                   &     $\sim$  1 10$^{-10}$      & $\sim$ 6 10$^{-10}$   \\
%HCN                 &    $\sim$   4 10$^{-11}$       & $\sim$ 2 10$^{-10}$   \\ 
%SiO                  &     $\sim$  7 10$^{-12}$        & $\sim$ 2 10$^{-12}$   \\  \hline 
SiO/C$^{34}$S  &    $\sim$ 2                           & $\sim$ 0.1 \\
CN/N$_2$H$^+$ & $\sim$ 1                             &  $\sim$ 6   \\
HCN/N$_2$H$^+$  & $\sim$ 0.4                                 & $\sim$ 2 \\
DCO$^+$/HCO$^+$  & $\sim$ 0.008  & $\sim$ 0.002  \\
D$_2$CO/DCO$^+$ &  $\sim$ 1     & $\sim$10   \\ \hline
\end{tabular}
\end{table}

\section{Summary and Conclusions: Chemical-clocks in intermediate-mass YSOs}
We have carried out a molecular survey towards the IM YSOs
FIRS~2 and LkH$\alpha$~234. Our survey confirms that 
protostellar envelopes are very complex objects composed by several
components characterized by different physical and chemical properties:
\begin{enumerate}
\item A cold envelope observationally 
characterized by narrow lines ($\Delta$v$\sim$1 km s$^{-1}$)
and low kinetic temperatures (T$_k$ $<$ 50 K). This component is well traced by
the species NH$_3$,N$_2$H$^+$, H$^{13}$CO$^+$, DCO$^+$ and N$_2$D$^+$.

\item A warm envelope observationally
characterized by large linewidths ($\Delta$v$>$3 km s$^{-1}$)
and kinetic temperatures (T$_k$$>$50 K). The low energy transitions of
 CS, C$^{34}$S, CH$_3$OH, and H$_2$CO arise in this warm component. 

\item A hot core characterized by high densities (n$>$10$^6$ cm$^{-3}$)
and a high kinetic temperature (T$_k$$>$100 k). The symmetric rotor CH$_3$CN
seems to be the best tracer of this hot component.

\item In addition to these envelope components, energetic outflows are associated with very YSOs. 
In addition to CO and its isotopes, the SiO emission is very likely the best tracer of the outflow component
for very young stellar objects.

\end{enumerate}

Once, we have used chemistry to determine the physical structure
of the YSOs, FIRS~2 and LkH$\alpha$~234, we can
determine  the evolution of the protostellar envelopes
of IM stars during the protostellar phase.
FIRS~2 is an IM Class 0 objects while LkH$\alpha$~234 is
a very young (and still deeply embedded) HBe star.
As expected, different physical conditions and chemistry are found in these objects.
The Class 0 IM is a cold object (T$_k$ $\sim$ 13 K) in which molecular depletion
is still important. 
We have no evidence of a change in the N$_2$H$^+$ abundance between both objects. 
The decrease in the N$_2$H$^+$ column density observed in Fig.~17 is the consequence of the
disruption of the cold envelope during the stellar evolution.
However, we have found evidences for H$^{13}$CO$^+$ depletion in FIRS~2.
Molecules like CH$_3$OH and H$_2$CO are expected to
trace mainly the warm region in which grain mantles have been evaporated. 
The column densities of these molecules remain constant (or increase) from the Class~0 IM to the more 
evolved HBe star, suggesting that the abundances of these molecules and/or the mass of the 
warm gas increases with the protostellar evolution. 
The detection of CH$_3$CN and the high temperature derived
from it (T$_k$$\sim$63 K) shows that a hot core has developed in the
Class~0 protostar FIRS~2. 
Thus, our results suggest an evolutionary
sequence in which as the protostar evolves to become a visible star,  the total column density of
gas is decreasing while the amount of warm gas remains quite constant or slightly increase. 
These physical and chemical changes imply important changes in the 
beam averaged column densities during the protostellar evolution.

Based on our observational study of FIRS 2 and LkH$\alpha$ 234, 
we propose some abundance ratios that can be used as chemical clocks 
in YSOs. These ratios are defined to be useful 
tools to discern between different evolutionary stages of YSOs, but  do
not correspond to the actual abundance ratios in any of the envelope components.
They have been calculated with beam averaged column densities and, as largely 
discussed in this paper, are the consequence of a complex physical and chemical 
evolution in the whole envelope.

In Table 10 we list the proposed chemical clocks. 
We find the maximum variation between the Class 0 and the HBe object when 
comparing the SiO/C$^{34}$S ratio.
This ratio decreases by a factor of $\sim$ 20 between these two objects. This
decrease is the consequence of the decay of the bipolar outflow phenomenum 
during the protostellar evolution. This ratio is specially useful to determine
the evolutionary stage of  the youngest objects
which are  associated with the most energetic bipolar outflows. 
The nitrogen chemistry is also useful to determine
the evolutionary stage of YSOs. The CN/N$_2$H$^+$
and HCN/N$_2$H$^+$ ratios are larger by a factor of $\sim$ 6 in
the HBe star than in the Class 0 object. 
As commented above this is mainly due to the fact that the LkH$\alpha$234
envelope is thinner and warmer  than that of the FIRS~2. 
These ratios would probably be more useful to differentiate between objects
in the late protostellar evolution when the protostellar envelope
becomes optically thinner. Finally, the deuterated
species could also be a good indicator of the protostellar evolution. 
The DCO$^+$/HCO$^+$ ratio decreases by a factor $\sim$ 4 due to the warmer 
envelope in LkH$\alpha$ 234. However, we can have a different behavior in
the deuterated species whose emission arise in the warm envelope component.
This is the case of the doubly deuterated compound D$_2$CO. We obtain the
largest contrast in the abundance ratio if we compare the D$_2$CO and DCO$^+$
abundances.
The D$_2$CO/DCO$^+$ ratio increases by a factor of 10 from the Class 0 to the
Herbig Be star. However, we should be cautious using this ratio because we are
comparing species arising in different regions of the envelope.

\acknowledgements
We are grateful to the MPIfR and IRAM staff in Pico de Veleta for their support during the
observations. This work has been partially supported
by the Spanish DGICYT under grant AYA2003-07584 and Spanish DGI/SEPCT under
grant ESP2003-04957. J.R.R. acknowledges the financial support from AYA2003-06473.

\bibliographystyle{aa}
{}
\end{document}